\documentclass[9pt]{mypaperformat}
\pdfoutput=1
\usepackage{stmaryrd}
\renewcommand{\vec}[1]{\boldsymbol{#1}}

\title{Mechanics and Variability of Cell Sheet Folding in the Embryonic
Inversion of \emph{Volvox}}

\author[1\authfn{1}]{Pierre A. Haas}
\author[1\authfn{1}]{Stephanie S. M. H. H\"ohn}
\author[1,2\authfn{2}]{Aurelia R. Honerkamp-Smith}
\author[1]{Julius B. Kirkegaard}
\author[1]{Raymond E. Goldstein}
\affil[1]{Department of Applied Mathematics and Theoretical Physics, University of Cambridge, Cambridge, UK}
\affil[2]{Department of Physics, Lehigh University, Bethlehem, PA, USA}

\corr{R.E.Goldstein@damtp.cam.ac.uk}{REG}

\contrib[\authfn{1}]{These authors contributed equally to this work}
\presentadd[\authfn{2}]{Department of Physics, Lehigh University, Bethlehem, PA, USA}

\begin{document}

\maketitle

\begin{abstract}
Many embryonic deformations during development are the global result of local cell shape changes and other local active cell sheet deformations. Morphogenesis does not only therefore rely on the ability of the tissue to produce these active deformations, but also on the ability to regulate them in such a way as to overcome the intrinsic variability of and geometric constraints on the tissue. Here, we explore the interplay of regulation and variability in the green alga \emph{Volvox}, whose spherical embryos turn themselves inside out to enable motility. Through a combination of light sheet microscopy and theoretical analysis, we quantify the variability of this inversion and analyse its mechanics in detail to show how shape variability arises from a combination of geometry, mechanics, and active regulation.
\end{abstract}

\section{Introduction}
Julian Huxley's pronouncement, ``In some colony like [the green alga] \emph{Volvox}, there once lay hidden the secret to the body and shape of [humans]'' \citep{huxley}, emphasises that morphogenesis across kingdoms relies on the fundamental ability of organisms to, firstly, produce active forces that drive the deformations of cell sheets underlying the development of many organs and tissues and, secondly, regulate these active deformations in such a way to complete morphogenesis. Unravelling the biomechanics of these processes is therefore of crucial importance to understand pathological errors and foster bioengineering to address these errors \citep{sasai12}. Local cellular changes can produce forces that are transmitted along the cell sheet to drive its global deformations \citep{lecuit07,lecuit11}. Simple events of cell sheet folding such as ventral furrow formation in \emph{Drosophila} can be driven primarily by cell shape changes \citep{sweeton91}. In more complex metazoan developmental processes such as gastrulation \citep{leptin05,wang09}, optic cup formation \citep{fuhrmann10,chauhan15}, neurulation \citep{lowery04,vija17} and related processes \citep{sherrard10}, the effect of such cell shape changes is overlaid by that of other cellular changes such as cell migration, cell intercalation, cell differentiation, and cell division. 

In all of these processes however, these local cellular changes occur in specific regions of the cell sheet and at specific stages of morphogenesis. On the one hand, the spatio-temporal distribution of these local cellular changes affects the global tissue shape. On the other hand, a certain amount of noise is unavoidable in biological systems; indeed, it may even be necessary for robust development, as demonstrated for example by \cite{hong16}, who showed that variability in cell growth is necessary for reproducible sepal size and shape in \emph{Arabidopsis}. 
While some processes may be subject to less intrinsic variability than others, one must therefore ask: how are these processes orchestrated so that development can complete despite the intrinsic biological variability? Differences in the observed shapes of organisms at certain stages of development (i.e. what one might term their geometric variability) stem from a combination of mechanical variability (i.e. differences in mechanical properties or mechanical state) and active variability (i.e. differences in the active forces generated by individual cells). What experimental data there are suggest that the mechanical properties are subject to a large amount of variability \citep[and references therein]{vondassow07}. Finally, differences in the mechanical stress state of the tissue are another facet of mechanical variability that is induced by active variability. 

The first mechanical models of morphogenesis \citep{odell81} represented cells as discrete collections of springs and dashpots; they were soon followed by elastic continuum models \citep{hardin86,hardin88}. Notable among this early modelling of morphogenesis is for example the work of \citeauthor{davidson95} (\textbf{\textit{\citeyear{davidson95}}}, \textbf{\textit{\citeyear{davidson99}}}), who combined models of several mechanisms of sea urchin gastrulation with measurements of mechanical properties to test the plausibility of these different mechanisms. These models heralded the emergence of a veritable plethora of mechanical modelling approaches over the subsequent decades \citep{fletcher17}, though the choice of model must ultimately be informed by the questions one seeks to answer \citep{rauzi13}. More recent endeavours were directed at deriving models that can represent the chemical and mechanical contributions to morphogenesis and their interactions \citep{howard11} and at establishing the continuum laws that govern these out-of-equilibrium processes \citep{prost15}. 

There is, however, a rather curious gap in the study of the variability of development: the importance of quantifying the morphogenesis and its variability has been recognised \citep{cooper08,oates09}, yet accounts of the variability of development, e.g. in the loach \citep{cherdantsev05,cherdantsev16}, have often been merely descriptive. For this reason, the interplay between mechanics and active variability has seemingly received little attention and hence a question we believe to be fundamental appears to lie in uncharted waters: how does active variability lead to geometric variability? Conversely, what does geometric variability tell us about active variability?

\begin{figure}
\centering\includegraphics{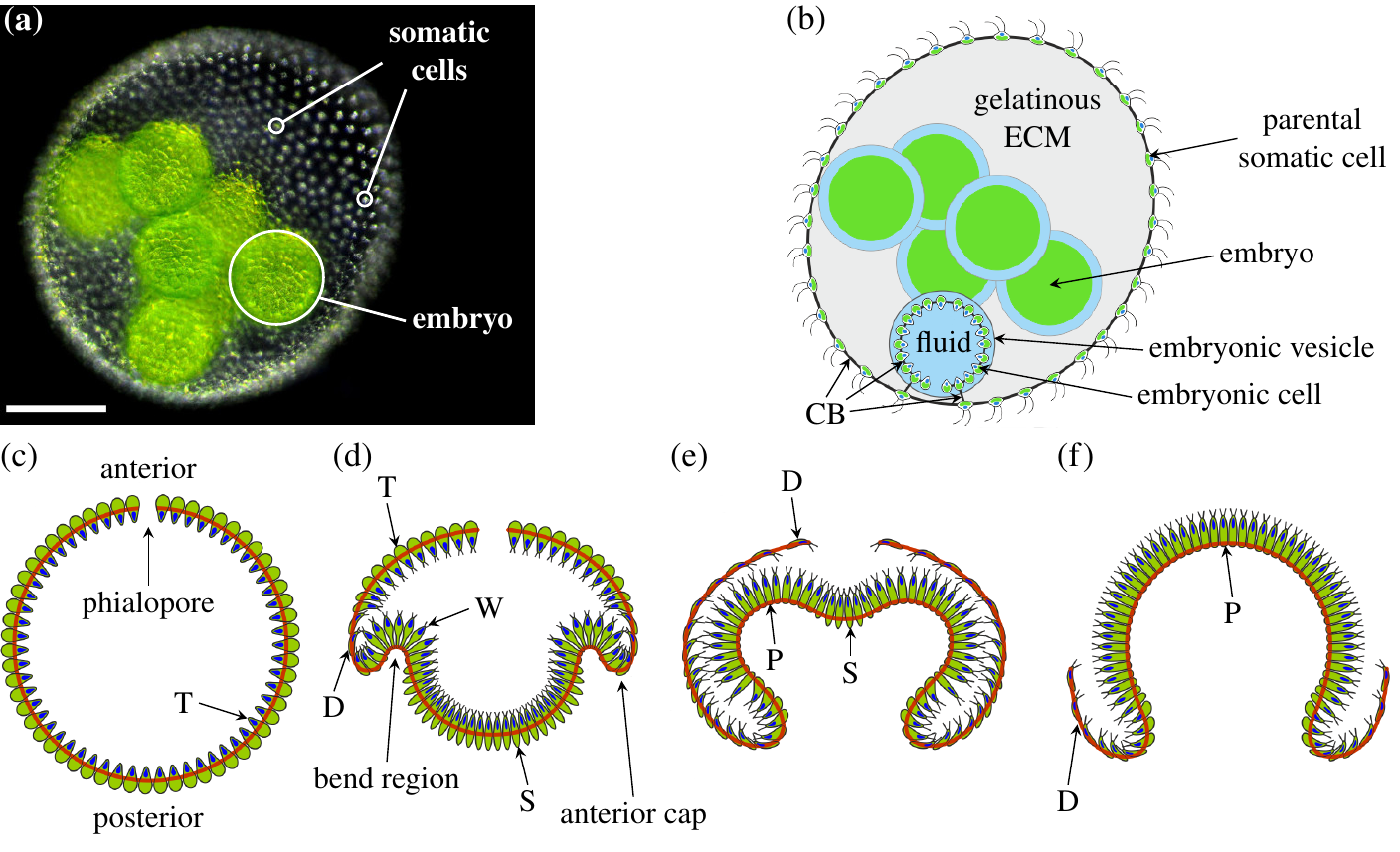}
\caption{Habitus of and embryonic inversion in \emph{Volvox globator}. (a)~Adult spheroid with somatic cells and one embryo labelled. Scale bar: $50\,\text{\textmu m}$. (b)~Schematic drawing of \emph{Volvox} globator parent spheroid with embryos. ECM: extracellular matrix. CB: cytoplasmic bridges. (c)~Schematic drawing of \emph{Volvox} embryo before inversion, with anterior and posterior poles, and phialpore labelled. Cells are teardrop-shaped [T]. (d)~\emph{Volvox} invagination: the formation of wedge-shaped cells [W] in the bend region initiates inversion. At the same time, cells in the posterior become spindle-shaped [S], while cells in the anterior close to the anterior cap become disc-shaped [D]. (d)~At the end of posterior inversion, cells in the whole of the anterior hemisphere are disc-shaped, while cells in the bend region are pencil-shaped [P]. (f)~As the anterior hemisphere peels over the inverted posterior, more and more cells become pencil-shaped. Red lines in panels (c--f) mark position of cytoplasmic bridges. Panels (c--f) adapted from \cite{hohn11}.}
\label{fig:volvox} 
\end{figure}

\begin{figure}
\begin{fullwidth}
\centering\includegraphics{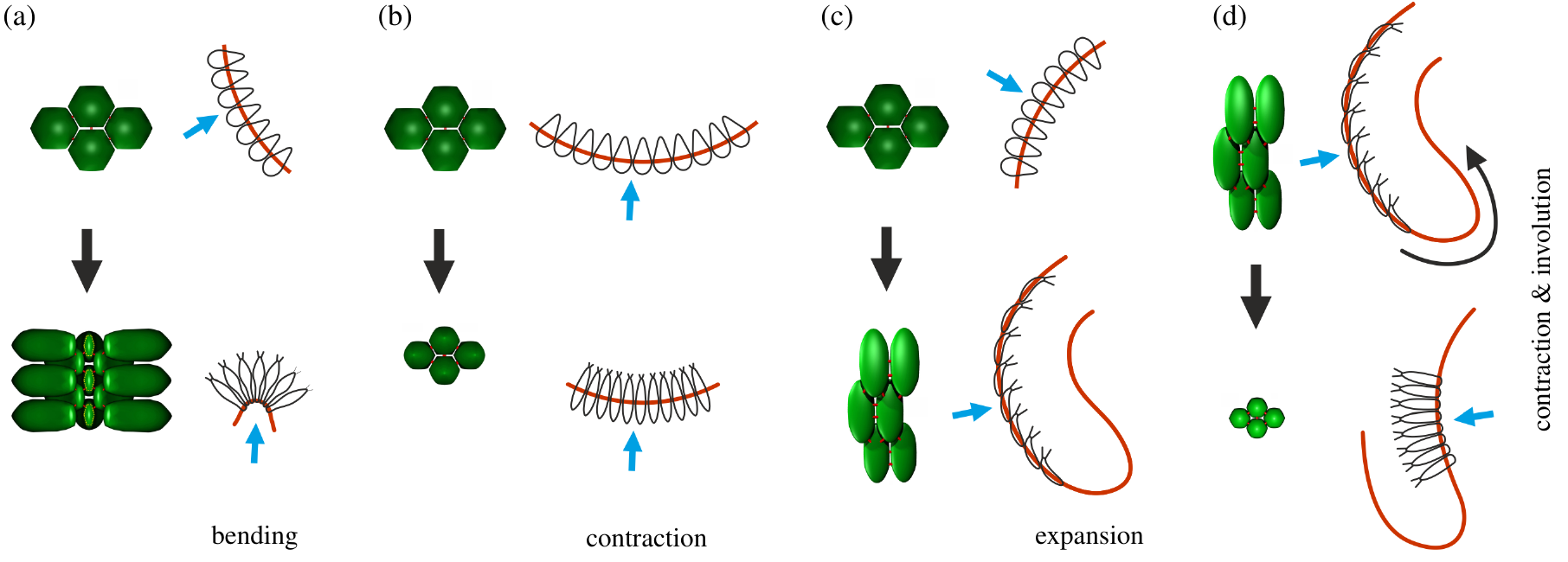}
\caption{Cell shape changes in \emph{Volvox globator} \citep{hohn11} associated with bending and stretching of the cell sheet. Cell shape changes (black arrows) (a)~from teardrop-shaped to paddle-shaped cells in combination with movement relative to the cytoplasmic bridges (CBs), associated with invagination of the bend region; (b) from teadrop-shaped to spindle-shaped cells, associated with contraction of the posterior hemisphere; (c) from teardrop-shaped to disc-shaped cells, associated with expansion of the anterior hemisphere (before opening of the phialopore) and (d) from disc to pencil-shaped, associated with contraction of the anterior hemisphere and involution over the anterior cap. Red line: position of the CBs, blue arrows: direction of view of cell groups shown.}
\label{fig:cellshapechanges} 
\figsupp{Cell movement relative to cytoplasmic bridges.}{Cell movement relative to cytoplasmic bridges. A motor protein, the kinesin InvA, is associated with cortical microtubules and an unknown structure within the cytoplasmic bridges in \emph{Volvox carteri} \citep{nishii03}. As the cells in the bend region develop think stalks, InvA `walks' towards the plus end of the microtubules, moving the cells until they are connected at the tips of their stalks.}{\includegraphics{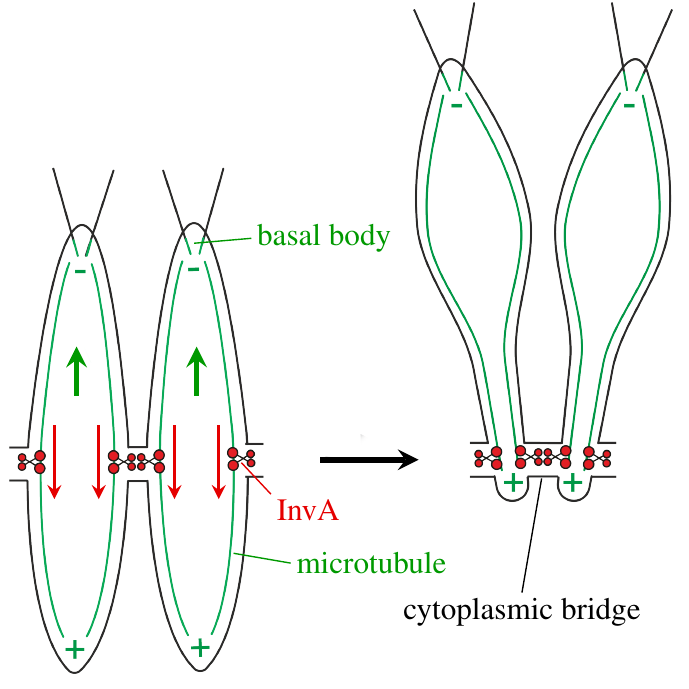}}
\end{fullwidth}
\end{figure}

This is the question that we explore in this paper in the context of the development of the multicellular green alga \emph{Volvox} (Fig.~\ref{fig:volvox}a). \emph{Volvox} and the related Volvocine algal genera have been recognised since the work of \cite{weismann} as model organisms for the evolution of multicellularity \citep{kirkbook,kirkessay,herron16}, spawning more recent investigations of kindred questions in fluid dynamics and biological physics \citep{goldstein15}. The cells of \emph{Volvox} (Fig.~\ref{fig:volvox}b) are differentiated into biflagellated somatic cells and a small number of germ cells, or gonidia, that will form daughter colonies \citep{kirkbook}. The somatic cells in the adult are embedded in a glycoprotein-rich extracellular matrix \citep{kirk86,hallmann03}. The germ cells undergo several rounds of cell division, after which each embryo consists of several thousand cells arrayed to form a thin spherical sheet confined to a fluid-filled vesicle. Cells are connected to their neighbours by cytoplasmic bridges (Fig.~\ref{fig:volvox}b), thin membrane tubes resulting from incomplete cell division \citep{green81a,green81b,hoops06}. Those cell poles whence will emanate the flagella however point into the sphere at this stage, and so the embryos must turn themselves inside out through an opening at the anterior pole of the cell sheet (the phialopore), to enable motility and thus complete their development \citep{kirkbook}. Because of this process of inversion, \emph{Volvox} has become a model organism for the study of cell sheet deformations, too~\citep{kirkchapter,kirkreview,matt16}.

Inversion in \emph{Volvox} \citep{viamontes77,hohn11} and in related species \citep{hallmann06,iida11,iida13,hohn16} results from cell shape changes only, 
without the complicating additional processes found in metazoan development discussed above.  This simplification 
facilitates the study of morphogenesis.
While different species of \emph{Volvox} have developed different ways of turning themselves inside out \citep{hallmann06}, here, we focus on the so-called type-B inversion arising, for example, in \emph{Volvox globator} \citep{zimmermann25,hallmann06,hohn11}. This shares features such as invagination and involution with developmental events in metazoans \citep{keller11,feroze15,czerniak16}. This inversion scenario is distinct from type-A inversion, in which four lips open at the anterior of the shell and peel back to achieve inversion \citep{viamontes77}. Type-B inversion begins with the appearance of a circular bend region at the equator of the embryo (Fig.~\ref{fig:volvox}c,d, Fig.~\ref{fig:cellshapechanges}a): cells there become wedge-shaped by developing narrow basal stalks \citep{hohn11}. At the same time, the cells move relative to the cytoplasmic bridges so as to be connected at their thin stalks, thus splaying the cells and bending and, eventually, invaginating the cell sheet \citep{hohn11}. \cite{nishii03} showed that inversion is arrested in the absence of analogous motion of cells relative to the cytoplasmic bridges in type-A inversion in \emph{Volvox carteri}. The relative motion results from a kinesin associated to the microtubule cytoskeleton (Fig.~\ref{fig:cellshapechanges}, figure supplement~1); orthologues of this kinesin are found throughout the Volvocine algae \citep{kirkessay}. After invagination, the posterior hemisphere moves into the anterior (Fig.~\ref{fig:volvox}e), the phialopore widens and the anterior hemisphere moves over the subjacent posterior (Fig.~\ref{fig:volvox}f) while `rolling' over a second circular bend region, the anterior cap \citep{hohn11}. Additional cell shape changes (Fig.~\ref{fig:volvox}d--f, Fig.~\ref{fig:cellshapechanges}b--d) in the anterior and posterior hemispheres are implicated in the relative contraction and expansion of either hemisphere with respect to the other \citep{hohn11}. This plethora of cell shape changes is possible as \emph{Volvox} cells do not have a cell wall \citep{kirkbook}.

In a previous study \citep{hohn15}, we combined light sheet microscopy and theory to analyse the early stages of inversion, showing that only a combination of active bending and active stretching (i.e. expansion or contraction) can account for the cell sheet deformations observed during invagination. The crucial role of active stretching was also highlighted by \cite{nishii99} who showed that type-A inversion in \emph{Volvox carteri} cannot complete if acto-moysin mediated contraction is inhibited chemically. We later analysed the mechanics of this competition between bending and stretching in more detail \citep{haas15}. Here, we analyse experimentally the variability of the shapes of inverting \emph{Volvox globator} at consecutive stages of inversion. We refine our theoretical model to capture later stages of the inversion process, and finally combine theory and experiment to untangle the geometric, mechanical, and active contributions to the observed spatial structure of the shape variations.

\begin{figure}
\centering\includegraphics[width=0.7\textwidth]{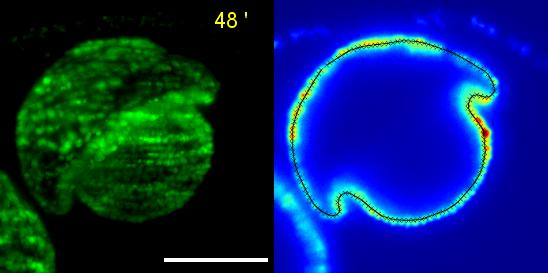}
\caption*{\textbf{Video 1.} Timelapse video of inverting \textit{Volvox globator} embryo from selective-plane illumination imaging of chlorophyll autofluorescence. Left: maximum intensity projection of z-stacks. Right: tracing of midsagittal cross-section (Methods). Scale bar: $50\,\text{\textmu m}$.}
\label{video:inversion}
\end{figure}

\section{Results}
We acquired three-dimensional time-lapse visualisations of inverting \emph{Volvox globator} embryos (Video~1) using a selective-plane-illumination-microscopy setup (Methods) based on the \textsc{OpenSPIM} system \citep{OpenSPIM}. Data were recorded for 13 parent spheroids containing, on average, 6 embryos. Summary statistics for 33 embryos were obtained from the recorded z-stacks and, for a more quantitative analysis of inversion, embryo outlines were traced on midsagittal sections of 11 of the recorded inversion processes, selected for optimal image quality (Methods).

\begin{figure}
\begin{fullwidth}
\centering\includegraphics{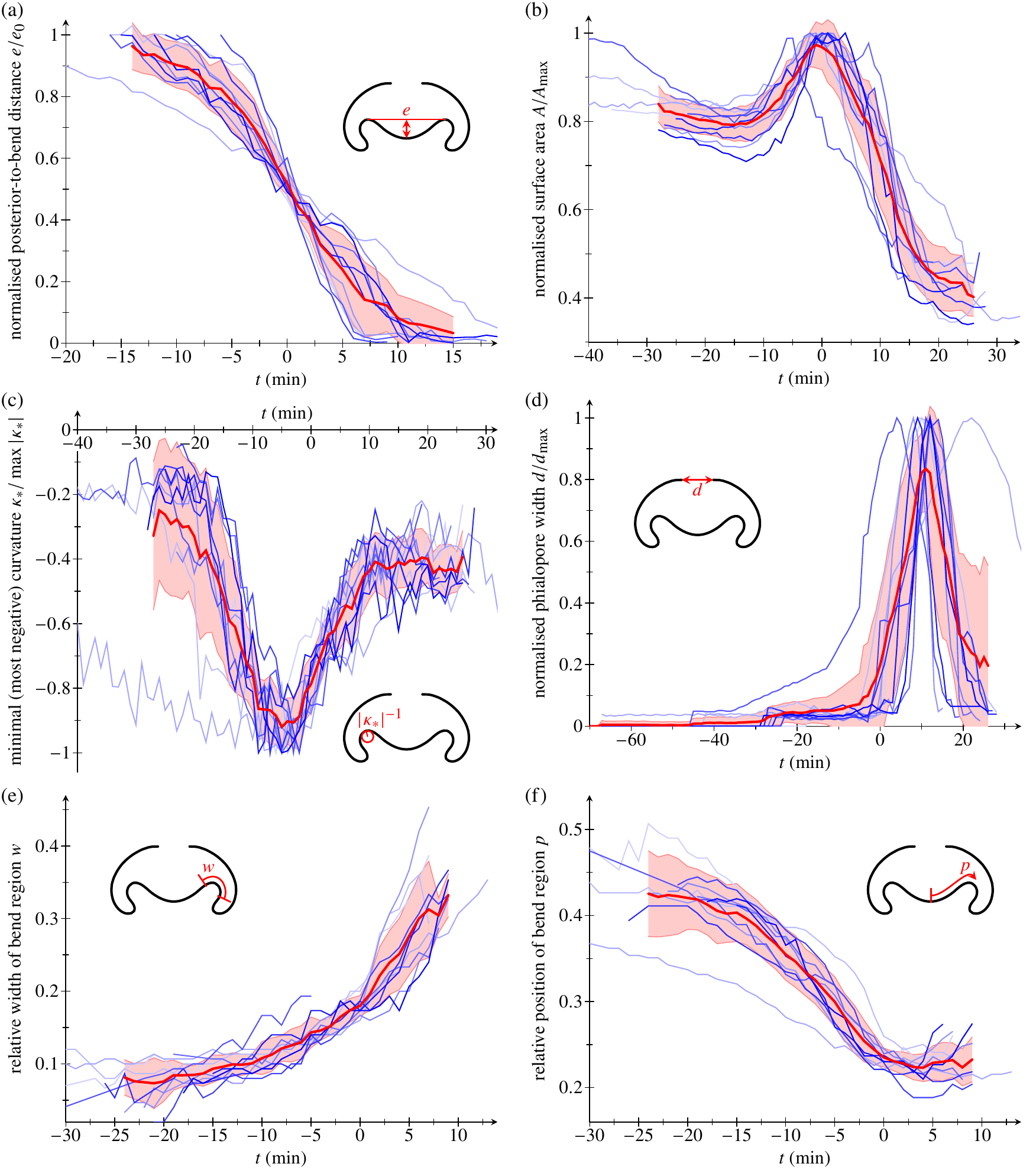}
\caption{Quantification of 11 inversion processes. (a)~Distance $e$ from the posterior pole to the plane of the circular bend region, normalised by its initial value $e_0$; (b)~Surface area $A$ computed from a surface of revolution obtained from each half of traced outlines, scaled by its maximal value~$A_{\max}$; (c)~Minimum (most negative) value $\kappa_*$ of the curvature in the bend region; (d)~Diameter of the phialopore $d$ normalised with its maximal value $d_{\max}$; (e)~Width $w$ of the bend region (region of negative curvature), normalised by arclength of the traced shape; (f)~Position $p$ of the bend region defined as the distance from the posterior pole to the centre of the bend region, normalised by arclength of the traced shape. Measurements (blue lines) on midsagittal embryo outlines are aligned at the time where $e$ reaches half of its initial value. Averages (red lines) and standard deviations thereof (red shaded areas) are shown for timepoints for which measurements were obtained for at least half of the quantified inversion processes. Insets: cartoons of definitions of $e$, $d$, $w$, $p$.}
\label{fig:measurements} 
\end{fullwidth}
\end{figure}

\begin{figure}
\centering\includegraphics{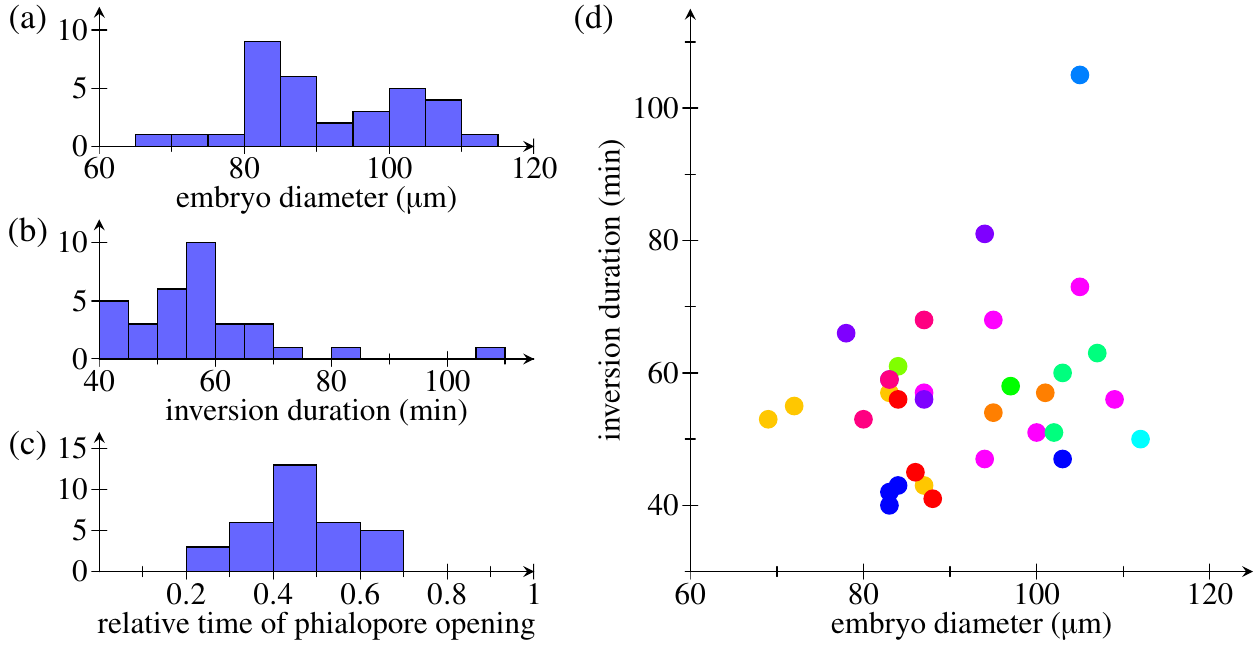}
\caption{Summary statistics for inversion from $N=33$ embryos. Histograms of (a)~duration of inversion, (b)~embryo diameter (after inversion), and (c)~relative time of phialopore opening. (d)~Duration of inversion plotted against embryo diameter (after inversion). Data points corresponding to embryos from the same parent spheroid are shown in the same colour.}
\label{fig:summarystats}
\figsupp{Characteristic points of measurements in Fig.~\ref{fig:measurements} and their variability.}{Timelines showing the variability of characteristic points of the measurements in Fig.~\ref{fig:measurements}, normalised by the total duration of inversion from appearance of the bend region until closure of the phialopore. a: first measurement of posterior-to-bend distance $e$ (when the tangent at the point of the most negative curvature $\kappa_\ast$ is horizontal); b: maximal negative curvature $\kappa_\ast$; c:~maximal surface area $A$; d: $e$ reaches half of its initial value; e: phialopore has widened to $20\%$ of its maximal diameter, f: $e$ reaches $10\%$ of its initial value; g: phialopore reaches its maximal diameter; h: phialopore has shrunk to $20\%$ of its maximal diameter. Posterior inversion (characteristic points a,d,f) is shown in red, and anterior inversion (characteristic points e,g,h) is shown in blue. The purple regions indicate an overlap of posterior and anterior inversion.}{\includegraphics{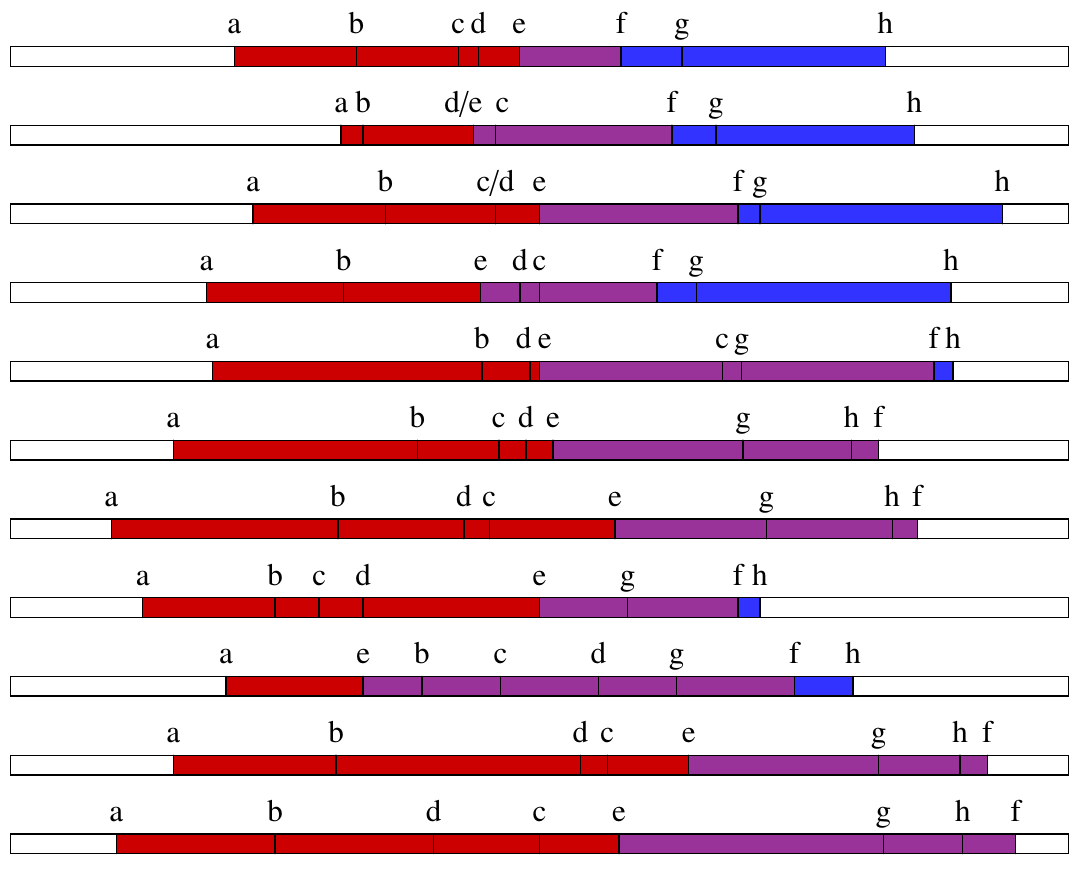}}
\end{figure}

In our previous work \cite{hohn15}, we discussed in detail three geometric descriptors of the traced embryo outlines, which we have reproduced for this dataset (Methods):
\begin{enumerate}[leftmargin=*]
 \item the distance $e$ (Fig.~\ref{fig:measurements}a) from the posterior pole to the plane of the circular bend region; this serves as an indicator of the progress of the `upwards' movement of the posterior hemisphere;
 \item the embryonic surface area $A$ (Fig.~\ref{fig:measurements}b), which was computed by determining a surface of revolution from each half of the midsagittal slice and averaging the two values for each timepoint;
 \item the minimal (most negative) value $\kappa_\ast$ of the meridional curvature in the bend region (Fig.~\ref{fig:measurements}c).
\end{enumerate}
We have computed three additional descriptors associated with the progress of later of inversion:
\begin{enumerate}[leftmargin=*,resume]
 \item the diameter $d$ of the phialopore (Fig.~\ref{fig:measurements}d) as an indicator of progress of inversion of the anterior hemisphere;
 \item the width $w$ of the bend region (Fig.~\ref{fig:measurements}e), where the bend region is defined as the region of negative curvature;
 \item the position of the bend region (Fig.~\ref{fig:measurements}f), measured along the arclength of the deformed shell from the posterior pole to the midpoint of the bend region.
\end{enumerate}
The computation of these descriptors is discussed in the Methods section. Each of these descriptors evolves in qualitatively similar ways in individual embryos, yet their evolution occurs over different timescales in different embryos and the local maxima in surface area (Fig.~\ref{fig:measurements}b) and in phialopore width (Fig.~\ref{fig:measurements}d) occurs at different relative times in different embryos. This initial impression of the variability of inversion is confirmed by the analysis of three summary statistics: (1)~the duration of inversion, from appearance of the bend region to closure of the phialopore, (2)~the diameter of the embryos post-inversion, (3)~the relative time during inversion at which the phialopore starts to open. Histograms of these quantities in Fig.~\ref{fig:summarystats}a--c reveal considerable variability, thus showing that the noise does not only affect the global duration of inversion, but also the relative timing of parts of it. We additionally note that there is no correlation between the size of an embryo and the duration of its inversion (Fig.~\ref{fig:summarystats}d), not even between embryos from the same parent spheroid.

It is natural to ask to what extent the different deformations of inversion must arise in a particular order: while invagination occurs before phialopore opening in all our samples, analysis of characteristic `checkpoints' of inversion (Fig.~\ref{fig:summarystats}, figure supplement 1) reveals that there is still considerable leeway in the timing of posterior inversion and phialopore opening. To further quantify the variability of inversion, we must define an average inversion sequence; our averaging approach must take into account these different types of variability.

\subsection{The Local Variability of Inversion}
\begin{figure}
\begin{fullwidth}
\includegraphics{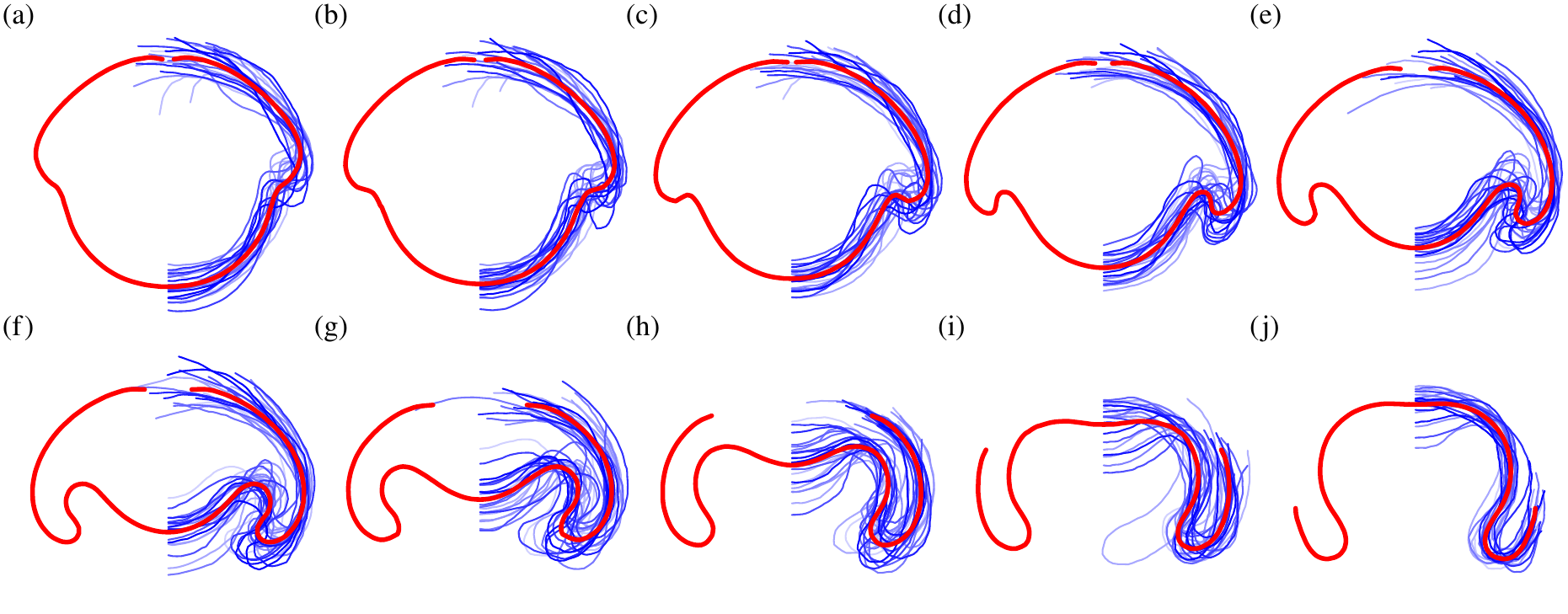}
\caption{Average Stages of Inversion. $N=22$ overlaid and scaled embryo halves from experimental data (lines in shades of blue), and averages thereof (red lines), for ten stages of inversion.}
\label{fig:averages} 
\figsupp{Geometry of averaging: alignment and local stretching.}{Averaging embryo shapes. (a)~Degrees of freedom for aligning shapes: after scaling and horizontal alignment by posterior pole (empty circles), only vertical alignment of shapes remains to be imposed by aligning centres of mass (filled circles). (b) Distributing points along arclength by averaging over different total arclengths ensures that the rims of the respective phialopores are matched up. Red line: average shape. (c) Distributing points along arclength at fixed distance between fitting points may yield a more faithful representation of part of the shape, but does not match up phialopores. Solid red line: average shape; dashed red line: average from (b) for comparison.}{\includegraphics{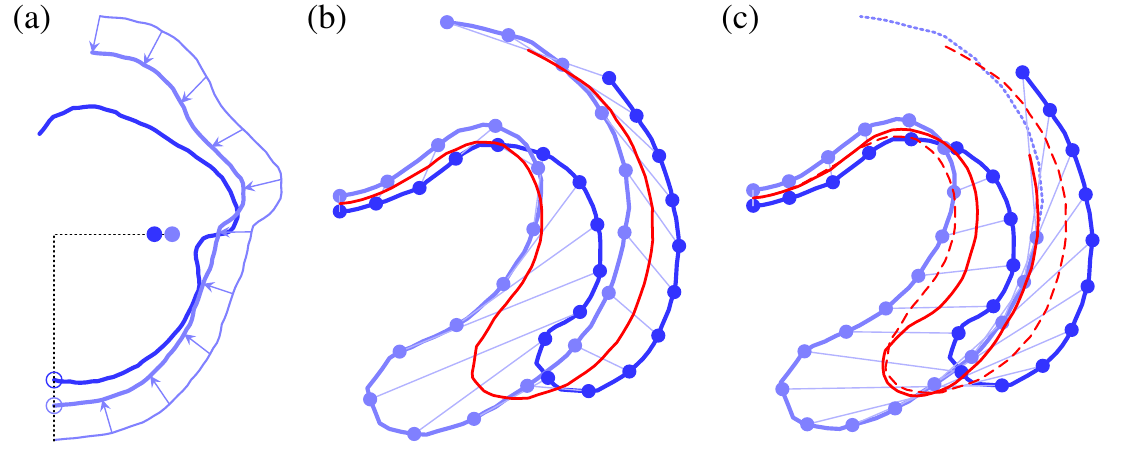}}
\figsuppfw{Alternative averaging approach 1: alignment by posterior-to-bend distance.}{Alternative averaging approach 1: alignment of embryos by the time-point where the posterior-to-bend distance $e$ reaches half of its initial value, without time stretching. $N=22$ overlaid and scaled embryo halves from experimental data (lines in shades of blue), and averages thereof (red lines), for ten stages of inversion. At late inversion stages, the average shapes are very noisy.}{\includegraphics{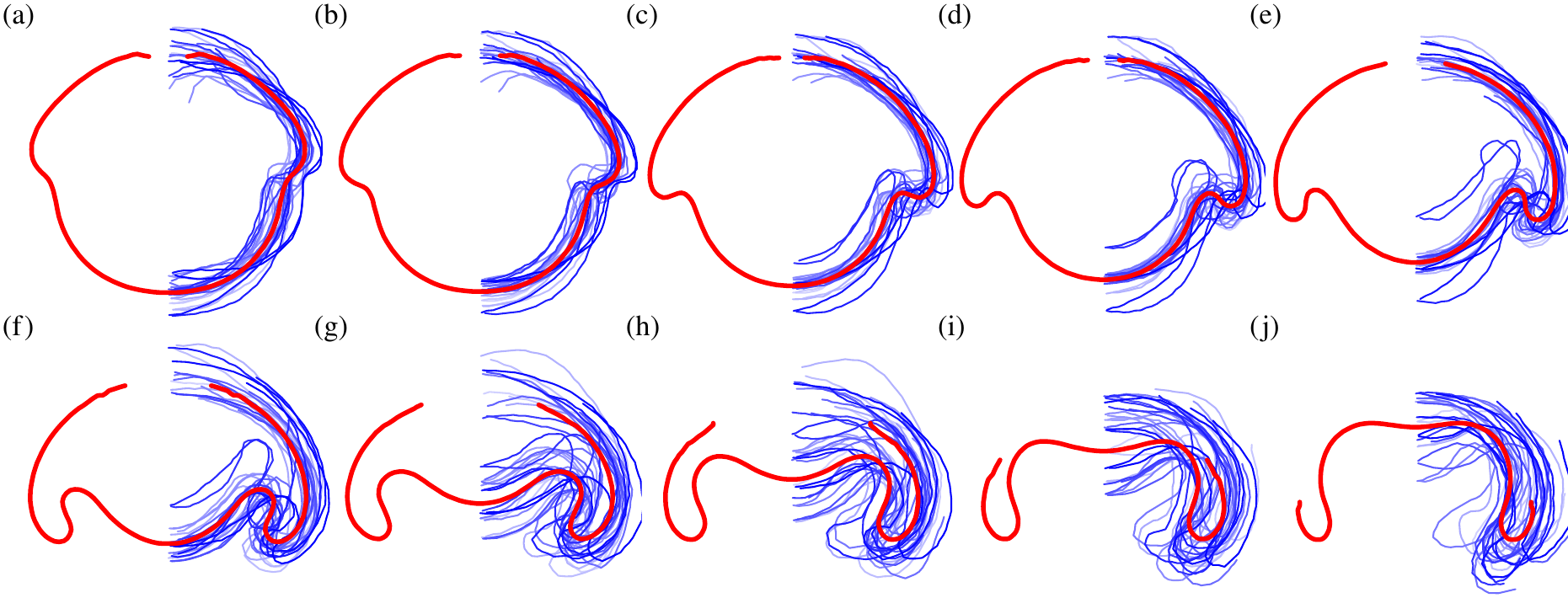}}
\figsuppfw{Alternative averaging approach 2: alignment with time variation, but without relative stretching of shapes.}{Alternative averaging approach 2: alignment of embryos with time stretching and with uniformly distributed averaging points (i.e. with only global scaling of embryos, without relative local stretching of embryo shapes). $N=22$ overlaid and scaled embryo halves from experimental data (lines in shades of blue), and averages thereof (red lines), for ten stages of inversion. Unsatisfactory `kinks' arise in the bend region.}{\includegraphics{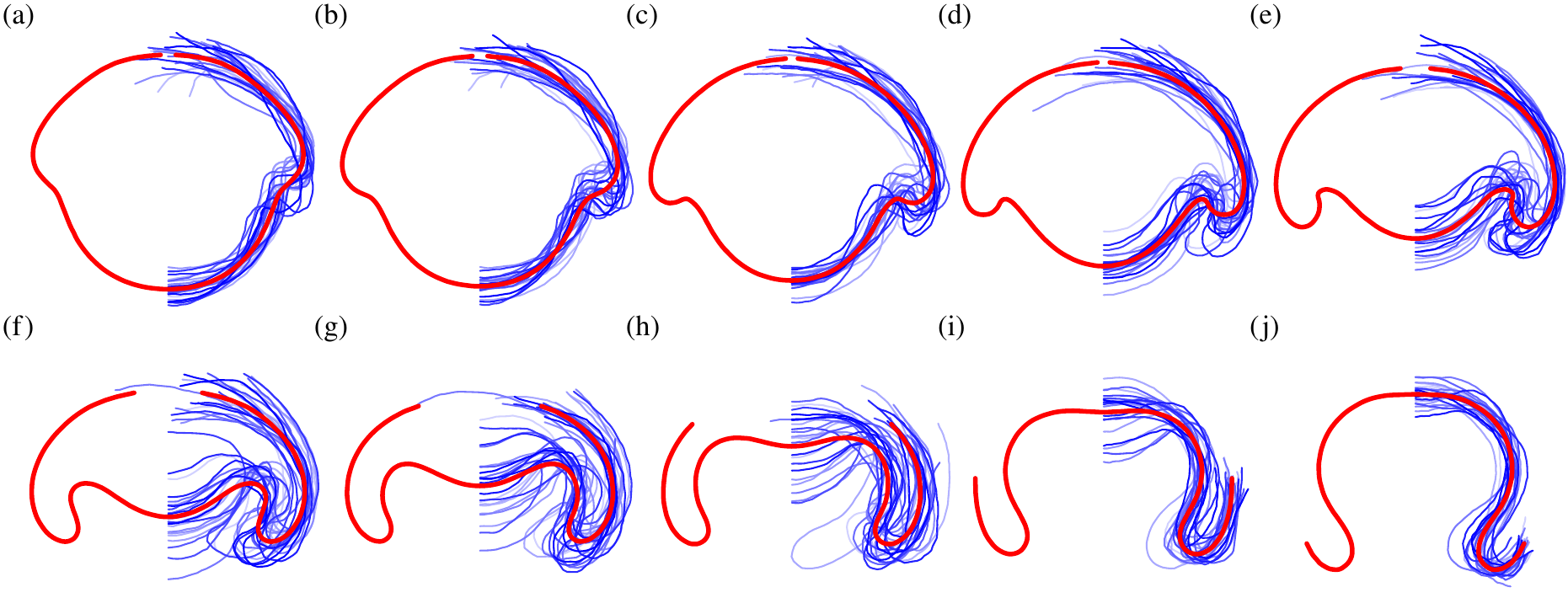}}
\end{fullwidth}
\end{figure}

To define an average inversion sequence and analyse its mechanics, we compare the local geometry of the traced curves. The question of how to define an appropriate metric for this kind of comparison goes back at least to the work of D'Arcy Thompson \citep{thompson}, and is altogether a rather philosophical one, to which there is no unique answer. Thompson showed for example how the outlines of fish of different species could be mapped onto one another by dilations, shears, and compositions thereof. In \emph{Volvox} inversion, these shape differences are likely to arise from variations in cell shape and variations in the positions of cell shape changes. Our averaging method must therefore allow for these local variations as well as for differences in the timing of the cell shape changes (as suggested by the analysis of the summary statistics), while recognising that the posterior poles and the rims of the phialopores of the different embryos must correspond to each other. Our approach is therefore based on minimising the Euclidean distance between individual embryo shapes and their averages, with alignments obtained using dynamic time warping (Methods and Fig.~\ref{fig:averages}, figure supplement 1). Results are shown in Fig.~\ref{fig:averages}.

Averaging approaches that do not consider both stretching in time of individual inversions and local stretching of corresponding points of individual shapes tend to give unsatisfactory results: the simplest averaging approach is to align the inversion sequences by a single time point, say when the posterior-to-bend distance reaches half of its initial value (Methods and Fig.~\ref{fig:averages}, figure supplement~2). The absence of time stretching, however, means that large variations arise at later stages of inversion. (Given the dramatic embryonic shape changes during inversion, it is not suprising that there should be no single parameter that could be used to align inversions of different embryos.) A better alignment is obtained if we allow stretching in time (Methods and Fig.~\ref{fig:averages}, figure supplement 3), but this method, without local stretching of individual shapes relative to each other, produces unsatisfactory kinks in the bend region of the average shapes (Fig.~\ref{fig:averages}, figure supplement 3).

\begin{figure}
\includegraphics{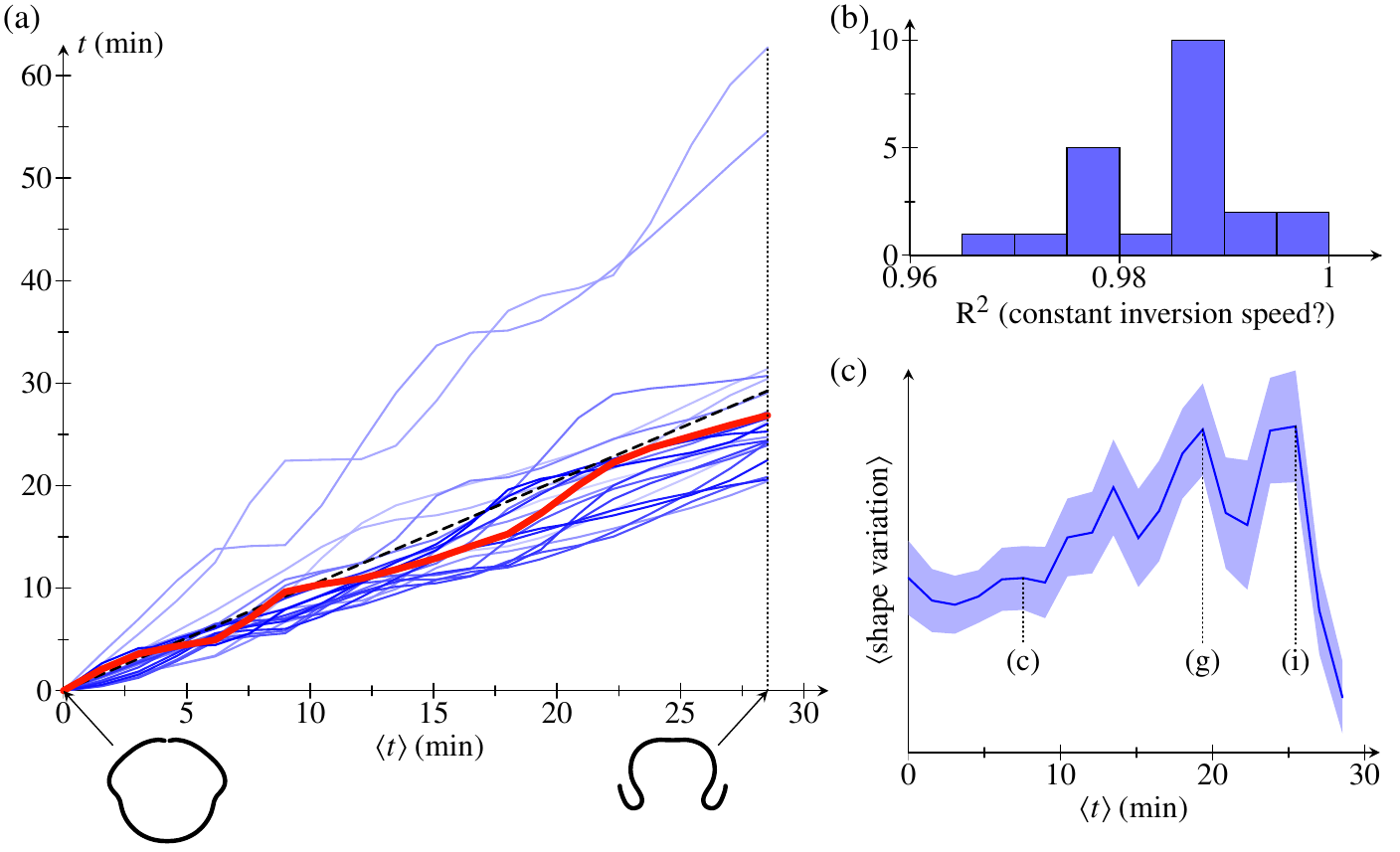}
\caption{Alignment Statistics. (a)~Timepoints $t$ for $N=22$ embryo halves (relative to first fitted timepoint) plotted against the mean values $\langle t\rangle$ of these times. Red line: time evolution illustrating non-linear progression of inversion. Insets: average embryo shapes at earliest and latest fitted times. (b) Histogram of $\mathrm{R}^2$ statistic for fits of the time evolutions in the first panel to a model of constant inversion speed. (c)~Mean shape variation (in arbitrary units), and standard errors thereof, against mean time $\langle t\rangle$. Corresponding panels in Fig.~\ref{fig:stdshapes} are marked for some data points.}
\label{fig:average_stats} 
\figsupp{Comparison of mean shape variation for different averaging methods.}{Mean shape variation against mean time $\langle t\rangle$, for the three averaging methods in Fig.~\ref{fig:averages}, showing that the averaging method using time stretching and local relative stretching of embryo shapes yields better averages than the two alternative averaging methods, especially at mid- to late-inversion stages. For alignment by posterior-to-bend distance, mean time was determined approximately by comparing the shapes in Fig.~\ref{fig:averages} and its figure supplements 2,3.}{\includegraphics{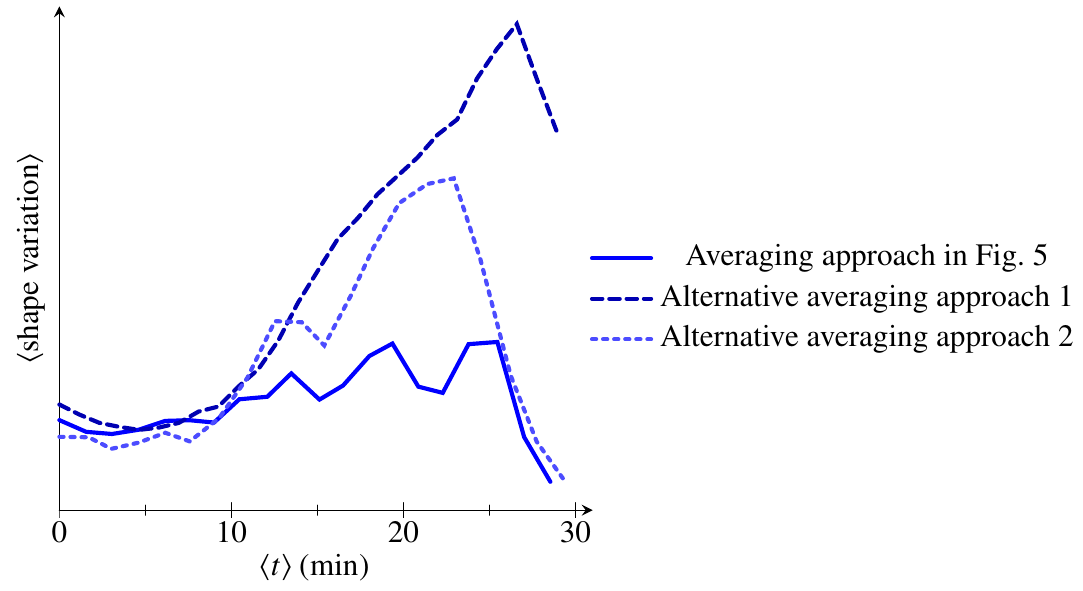}}
\end{figure}

The averages reveal that inversion seems to proceed at an approximately constant speed relative to the average inversion sequence (Fig.~\ref{fig:average_stats}a,b). However, the alignment reveals that different stages of inversion take different times in different embryos (Fig.~\ref{fig:average_stats}a), with some embryos seeming to linger in certain stages. This is the same non-linearity that we already saw earlier on the timelines in Fig.~\ref{fig:summarystats}, figure supplement 1 obtained from the measurements in Fig.~\ref{fig:measurements}.
 
\begin{figure}
\begin{fullwidth}
\includegraphics{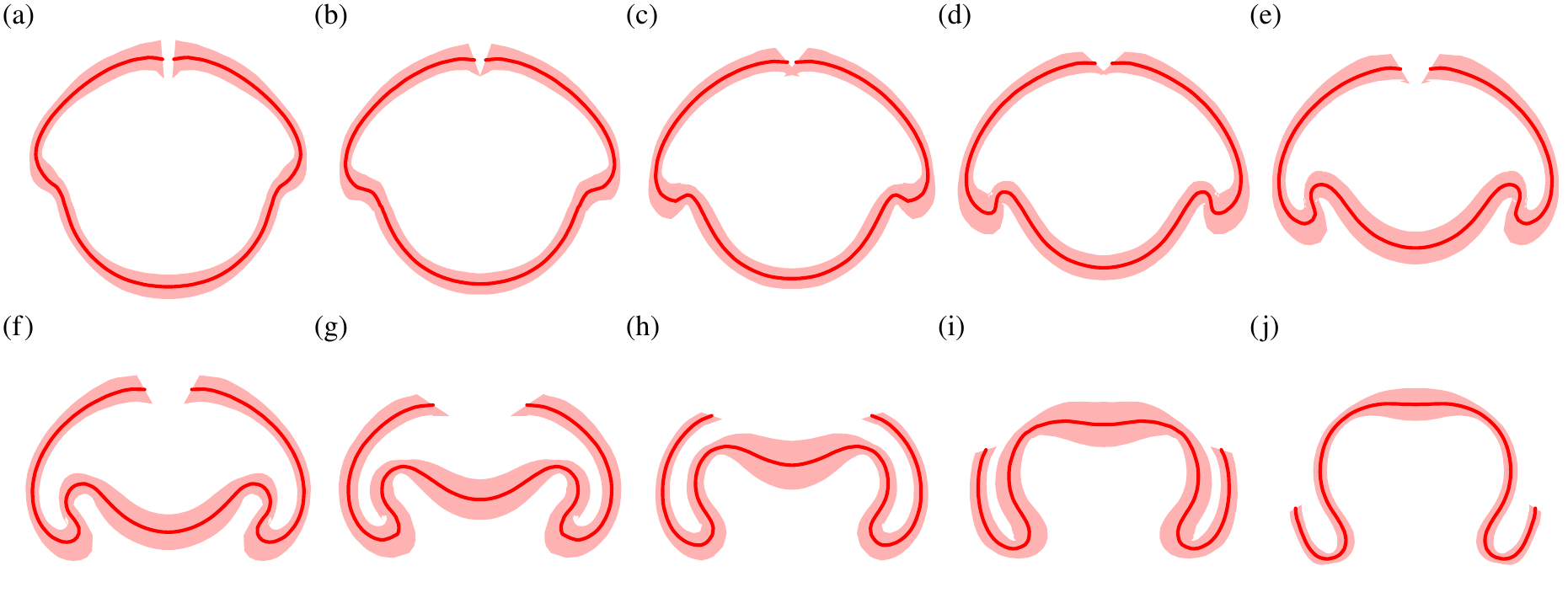}
\caption{Local Variations of \emph{Volvox} Shapes during Inversion. Average shapes from Fig.~\ref{fig:averages} (red lines) and corresponding standard deviation shapes (shaded areas).}
\label{fig:stdshapes} 
\end{fullwidth}
\end{figure}

To analyse the local variations of the embryo shapes, we define, at each point of the average shapes, a covariance ellipse. The curves that are parallel to the average shape and tangent to the covariance ellipse define what we shall term the standard deviation shape. These standard deviation shapes measure the variability of the average shapes and are shown in Fig.~\ref{fig:stdshapes}. The variations they represent naturally divide into two components: first, those variations that are parallel to the average shape, and second, those perpendicular to the average shape. The former represent mere local stretches of the average shapes, while the latter correspond to actual variations of the shapes; we shall therefore refer to the thickness of the standard deviation shapes as `shape variation' in what follows. We report the mean shape variation and its standard error in Fig.~\ref{fig:average_stats}c. This plot shows that the mean shape variation reaches a maximal value around the stages in Fig.~\ref{fig:stdshapes}g--i: different embryos start from the same shape and reach the same inverted shape after inversion (up to a scaling), but may take different inversion paths. Plotting the mean shape variation for different averaging methods (Fig.~\ref{fig:average_stats}, figure supplement 1), we confirm that the present averaging method yields a better alignment than the alternative methods discussed earlier. 

It is intriguing, however, to note the spatial structure of the local shape variations. In particular, during the early stages of posterior inversion (Fig.~\ref{fig:stdshapes}d--f), the shape variation is smaller in the active bend region than in the adjacent anterior cap (Fig.~\ref{fig:volvox}e, the second bend region of increased positive curvature). As the phialopore opens and the anterior begins to peel back over the partially inverted posterior (Fig.~\ref{fig:stdshapes}h) the relative shape variations becomes smaller in the anterior cap. The initially small variation in the bend region is especially intriguing since this is where cells become wedge-shaped to drive invagination, while the anterior cap bends passively \citep{hohn15}. In other words, the shape variation is reduced in the part of the cell sheet where the active cell shape changes driving inversion arise. This correspondence characterises what one might term, from a teleological point of view, a `good' inversion. We shall focus on a less exalted question, the answer to which will be falsifiable, however: how is this spatial structure of the variability related to the mechanics of inversion? Before addressing this question, we need to analyse the mechanics of inversion in some more detail.

\subsection{Active Bending and Stretching during Inversion}
Which active deformations are required for inversion? In our previous work~\citep{hohn15,haas15}, we addressed 
this question for the early stages of inversion: at a mechanical level of description, invagination arises from an interplay of active bending and stretching~\citep{hohn15,haas15} associated with different types of cell shape changes \mbox{(Fig.~\ref{fig:cellshapechanges}a--c)}. A key role is played by the cells close to the equator of the cell sheet (Fig.~\ref{fig:volvox}d, Fig.~\ref{fig:cellshapechanges}a), which become wedge-shaped~\citep{hohn11}, thus splaying the cells and hence imparting intrinsic curvature to the cell sheet~\citep{haas15}. Yet no such cell wedging has been reported at the anterior cap at later stages of inversion, when the anterior hemisphere peels back over the partly inverted posterior (Fig.~\ref{fig:volvox}f, Fig.~\ref{fig:cellshapechanges}d). 

To resolve this conundrum, we ask whether the additional cell shape changes observed during type-B inversion~\citep{hohn11} are sufficient to explain anterior peeling: cells in the anterior hemisphere have flattened, ellipsoidal shapes, while cells on the posterior side of the anterior cap are pencil-shaped (Fig.~\ref{fig:volvox}e, Fig.~\ref{fig:cellshapechanges}d). We have previously described the early stages of inversion using a mathematical model~\citep{hohn15} in which cell shape changes appear as local variations of the intrinsic (meridional and circumferential) curvatures $\kappa_s^0,\kappa_\phi^0$ and stretches $f_s^0,f_\phi^0$ of an elastic shell. We recall the difference between open, one-dimensional elastic filaments and two-dimensional elastic shells in this context: the former can simply adopt a shape in which the curvature and stretch are everywhere equal to their intrinsic values. For the latter, by contrast, the intrinsic curvatures and stretches may not be compatible with the global geometry, causing the shell to deform elastically and adopt actual (meridional and circumferential) curvatures $\kappa_s,\kappa_\phi$ and stretches $f_s,f_\phi$ different from the imposed intrinsic curvatures and stretches. In order to address these later stages of inversion, we must first generalise our previous mathematical model using ideas from morphoelasticity (Methods).
Indeed, that model was derived under the assumption of small strains. While the \emph{elastic} strains are small indeed (since the metric tensor, which describes the deformed shape, is close to the intrinsic tensor defined by the cell shape changes), the \emph{geometric} strains are large: both the metric tensor of the deformed shell and the intrinsic tensor differ considerably from the metric tensor of the undeformed sphere.

\begin{figure}
\centering\includegraphics{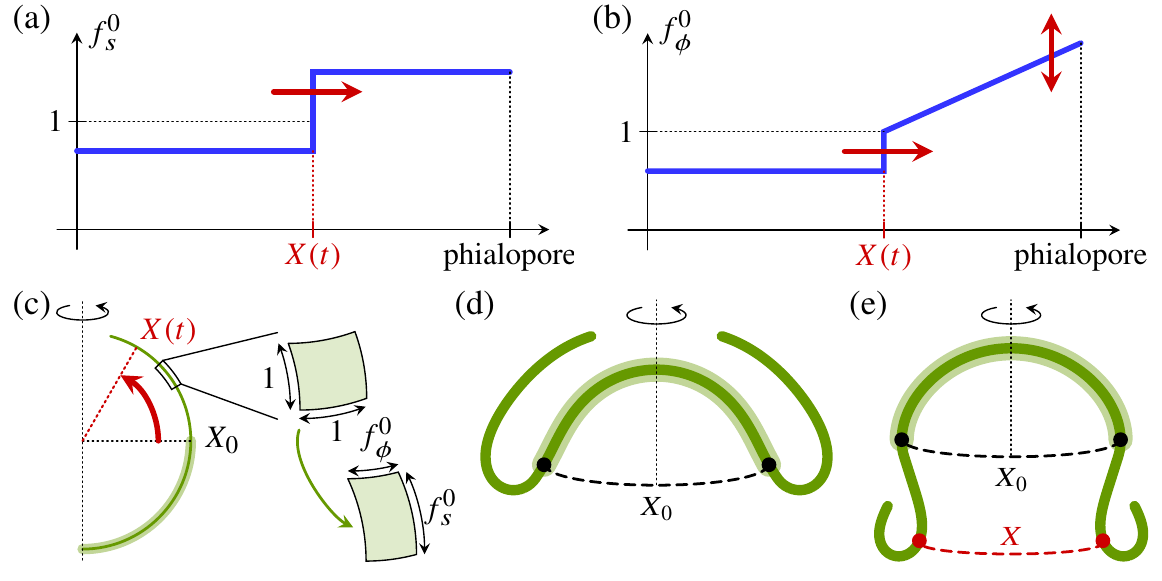}
\caption{Mechanics of Anterior Peeling. Functional form of (a) the meridional intrinsic stretch $f_s^0$ and (b)~the circumferential intrinsic stretch $\smash{f_\phi^0}$, with position $X(t)$ of the peeling front indicated. (c)~Definition of the position $X$ of the peeling front and its initial value $X_0$ at the equator of the undeformed shell. The shaded area indicates the posterior hemisphere in which the intrinsic curvatures are equal and opposite to those of the undeformed sphere. Inset: definitions of the intrinsic stretches $\smash{f_s^0,f_\phi^0}$. (d)~Shape before peeling, with inverted posterior hemisphere. (e)~Resulting shape after anterior peeling, just before phialopore closure, with $X_0$ and $X$ indicated.}
\label{fig:antpeeling} 
\end{figure}

To test whether anterior peeling can be achieved by contraction of the cell sheet alone, we impose functional forms for the intrinsic stretches of the shell (Fig.~\ref{fig:antpeeling}a--c) representing these cell shape changes, but we do not modify the intrinsic curvatures in the anterior hemisphere (Fig.~\ref{fig:antpeeling}c). In particular, the linear variation of the circumferential stretch in the anterior hemisphere represents the different orientations of the ellipsoidal cells at the phialopore (Fig.~\ref{fig:cellshapechanges}c), where the long axis is the circumferential axis, and at the anterior cap, where the long axis is the meridional axis \citep{hohn11}. In our quasi-static simulation, we approximate the shape in Fig.~\ref{fig:averages}h by a configuration with inverted posterior hemisphere (Fig.~\ref{fig:antpeeling}d), and displace the intrinsic `peeling front' (Fig.~\ref{fig:antpeeling}a,b). The shell responds by peeling (Fig.~\ref{fig:antpeeling}e), with shapes in qualitative agreement with the experimentally observed shapes. Since the peeling front is located at the anterior cap, where the shape variation is reduced during anterior peeling as discussed previously, we again see a correlation between reduced shape variations and the location of the active cell shape changes driving inversion. 

These considerations suggest that contraction is sufficient to drive the peeling stage of inversion, even without changes in intrinsic curvature. Although the position of the cytoplasmic bridges (Fig.~\ref{fig:cellshapechanges}d), on the inside end of the cells at the end of inversion~\citep{hohn11}, suggests that the intrinsic curvature may change sign in the anterior hemisphere, too, this appears to be a seconday effect. Hence intrinsic bending complements intrinsic stretching. By contrast, our previous work~\citep{hohn15} revealed that stretching complements bending during invagination. The roles of stretching and bending are thus interchanged during inversion of the posterior and anterior hemispheres, and the embryo uses these two different deformation modes for different tasks during inversion.

\subsubsection{Analysis of Cell Shape Changes}
\begin{figure}
\begin{fullwidth}
\includegraphics{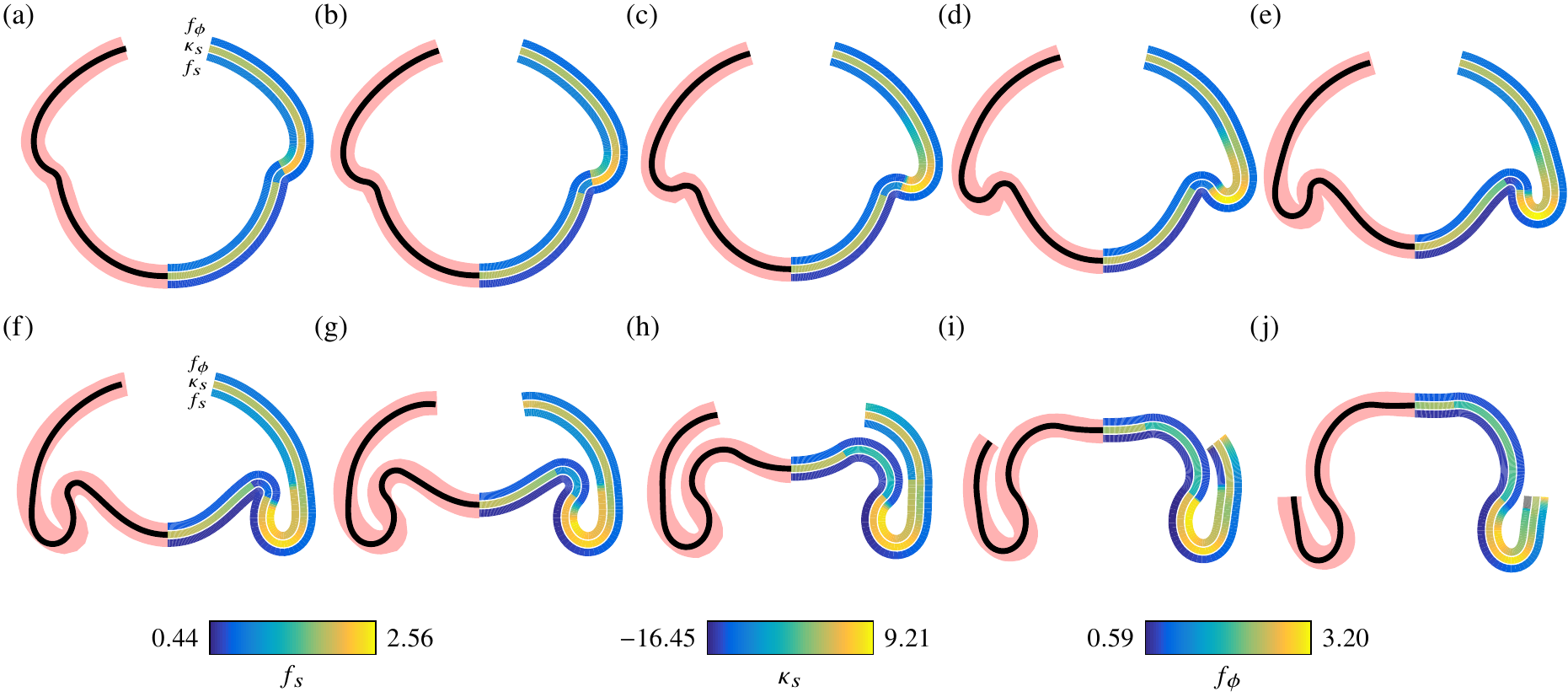}
\caption{Average Embryo Shapes reproduced by the elastic model. In each panel, the left half shows average shapes from Fig.~\ref{fig:averages} (thick red line) and corresponding fits (black line) from the elastic model for different stages of inversion. The right half shows colour-coded representations of the meridional curvature $\kappa_s$ and stretches $f_s$ and $f_\phi$ in the fitted shapes.}
\label{fig:fittedshapes} 
\figsupp{Geometric descriptors of average and fitted shapes.}{Geometric descriptors, as in Fig.~\ref{fig:measurements}, for the average shapes in Fig.~\ref{fig:averages} (red marks) and the fits of Fig.~\ref{fig:fittedshapes} (black line). (a)~Posterior-to-bend-distance $e$; (b)~Surface area $A$; (c)~Minimum (most negative) value $\kappa_\ast$ of the curvature in the bend region; (d)~Phialopore diameter $d$; (e)~Bend region width $w$; (f)~Bend region position $p$. The geometric descriptors are normalised as in Fig.~\ref{fig:measurements}; insets provide cartoons of definitions.}{\includegraphics{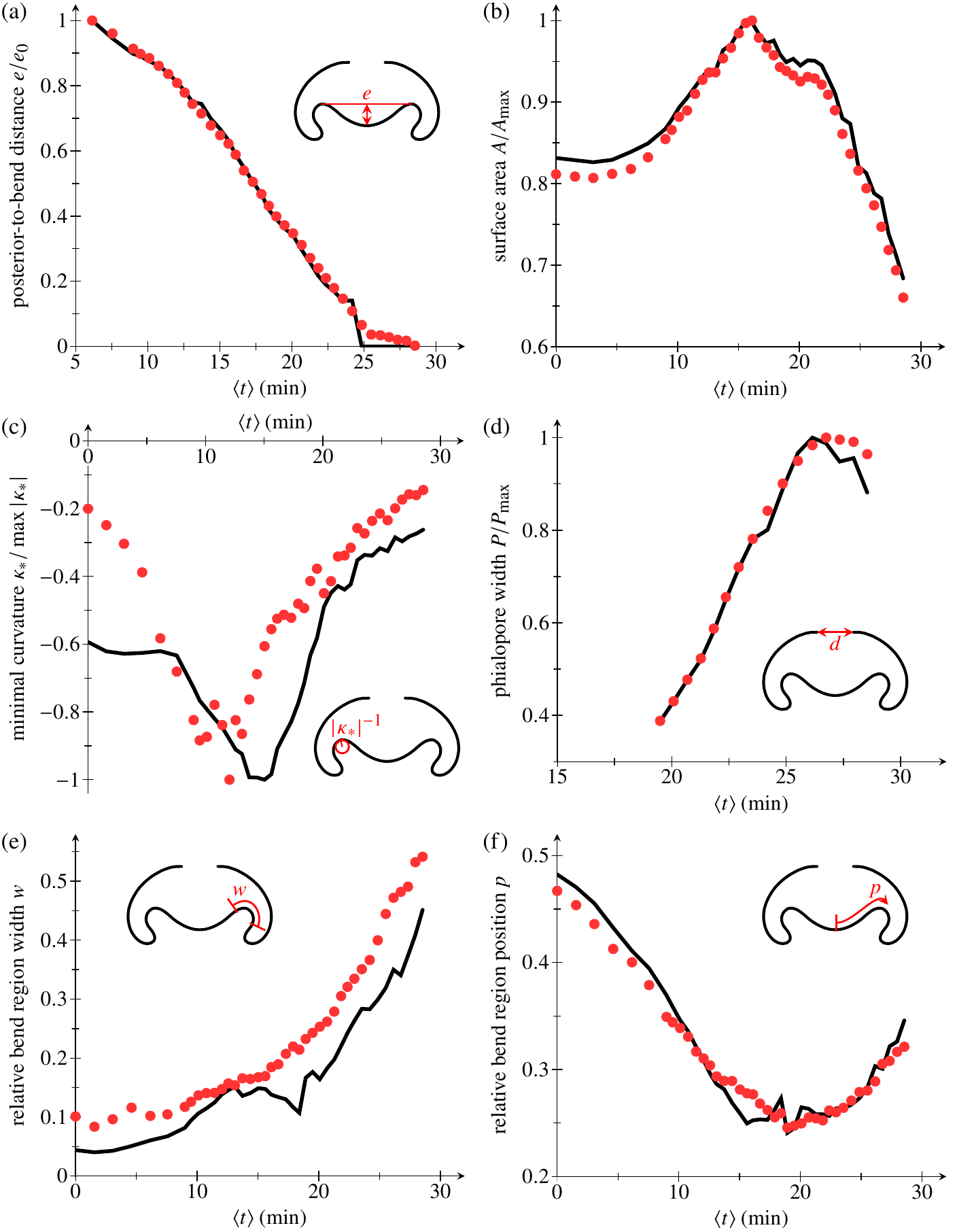}}
\end{fullwidth}
\end{figure}

For a more quantitative analysis of the data and to validate our model, we proceed to fit the elastic model to the experimental average shapes (Methods). In the model, we impose a larger extent of the phialopore than in the biological system, where the phialopore is initially very small (Fig.~\ref{fig:averages}a). This is an important simplification to deal with the discrete nature of the few cells that meet up at the phialopore. Nonetheless, using fifteen fitting parameters to represent previously observed cell shape changes~\citep{hohn11} in terms of the intrinsic stretches and curvatures (Methods), the model captures the various stages of inversion (Fig.~\ref{fig:fittedshapes}). This supports our interpretation of the observed cell shape changes (Fig.~\ref{fig:cellshapechanges}) and their functions. Comparing the geometric descriptors discussed previously (Fig.~\ref{fig:measurements}) for the experimental averages and the fitted shapes (Fig.~\ref{fig:fittedshapes}, figure supplement 1), we notice that the fitted shapes underestimate the width of the bend region. Because curvature is a second derivative of shape, it is not surprising that larger differences arise in the minimal bend region curvature of the average and fitted shapes (Fig.~\ref{fig:fittedshapes}, figure supplement 1).

\begin{figure}
\centering\includegraphics{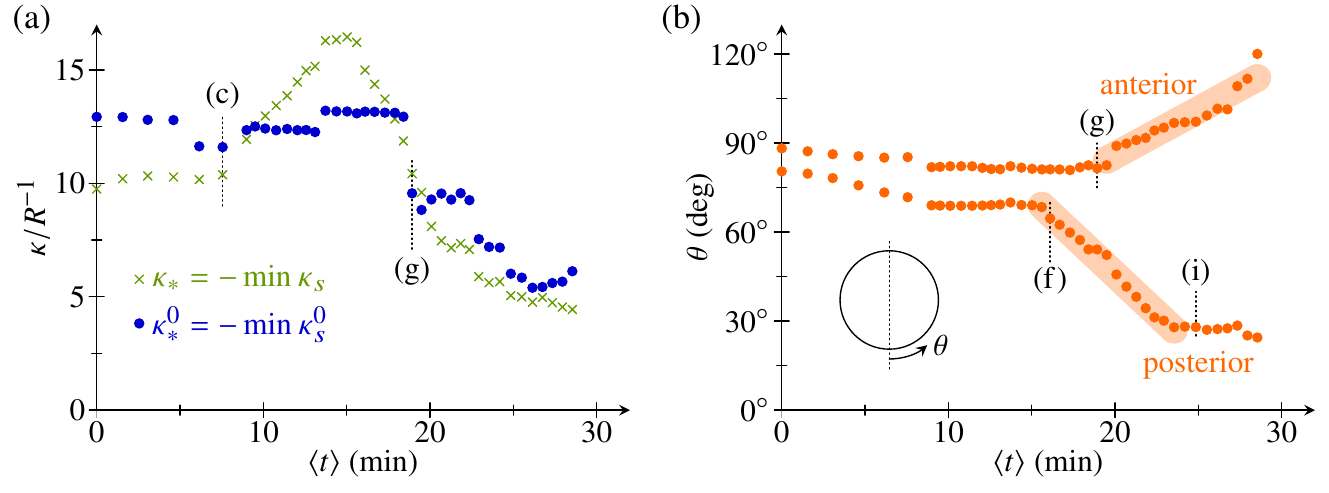}
\caption{Analysis of fitted parameters. (a) Plot of most negative values of the intrinsic and actual meridional curvatures, $\kappa^0_\ast=-\min{\kappa_s^0}$ and $\kappa_\ast=-\min{\kappa_s}$ against mean time $\langle t\rangle$. (b)~Positions of posterior and anterior limits of the bend region relative to the undeformed sphere, plotted against mean time $\langle t\rangle$. Thick lines indicate straight line fits. Corresponding panels in Fig.~\ref{fig:fittedshapes} are marked for some data points.}
\label{fig:params} 
\end{figure}

Nonetheless, the fitted values of the intrinsic curvature of the cell sheet also resolve a cell shape conundrum: during invagination, the curvature in the bend region increases (Fig.~\ref{fig:fittedshapes}), yet \cite{hohn11} reported similar wedge-shaped cells in the bend region at early and late invagination stages, although the number of wedge-shaped cells in the bend region increases as invagination progresses~\citep{hohn11}. The fitted parameters indeed suggest a constant value of the intrinsic curvature at early stages of inversion, while the actual curvature in the bend region increases (Fig.~\ref{fig:params}a). This serves to illustrate that the intrinsic parameters cannot simply be read off the deformed shapes and confirms that there is but a single type of cell change, expanding in a wave to encompass more cells, and thus driving invagination. It is only at later stages of inversion, when the wedge-shaped cells in the bend region become pencil-shaped~\citep{hohn11}, that both the intrinsic curvature and the actual curvature in the bend region decrease (Fig.~\ref{fig:params}a). The fitted shapes also yield the posterior and anterior limits of the bend region (Fig.~\ref{fig:params}b), i.e. the original positions, relative to the undeformed sphere, of the corresponding cells. Because of the varying spatial stretches of the shell, these positions cannot simply be read off the deformed shapes, but must be inferred from the fits. The fitted data suggest that invagination results from an intrinsic bend region of constant width, complemented by other cell shape changes (Fig.~\ref{fig:volvox}d, Fig.~\ref{fig:cellshapechanges}b,c). The 
region of wedge-shaped cells (and, by implication, of negative intrinsic curvature) starts to expand into the posterior at constant speed (i.e. at a constant number of cell shape changes per unit of time) between the stages in Fig.~\ref{fig:fittedshapes}e,f. Anterior inversion starts about five minutes later when this region begins to expand into the anterior just after the stage in Fig.~\ref{fig:fittedshapes}g. 

Fig.~\ref{fig:fittedshapes} also shows the stretches $f_s,f_\phi$ in the fitted shapes. It is particularly interesting to relate the values of $f_s,f_\phi$ in the fitted shapes to the measurements of individual cells by \cite{hohn11}: before inversion starts, the cells are teardrop-shaped, and measure $3-5\,\text{\textmu m}$ in the plane of the cell sheet. As invagination starts, the cells in the posterior hemisphere become spindle-shaped, measuring $2-3\,\text{\textmu m}$. This suggests values $f_s,f_\phi\approx 0.6-0.66$ in the posterior hemisphere during invagination, in agreement with the fitted data (Fig.~\ref{fig:fittedshapes}d). At later stages of inversion, the cells in the bend region become pencil-shaped, measuring $1.5-2\,\text{\textmu m}$ in the meridional direction, suggesting smaller values $f_s\approx 0.4-0.5$ there, again in agreement with the fitted data (Fig.~\ref{fig:fittedshapes}h). The large stretches $f_s>2$ seen in the anterior cap during inversion of the posterior hemisphere (Fig.~\ref{fig:fittedshapes}f) cannot be accounted for by the disc-shaped cells in the anterior (which only measure $4-6\,\text{\textmu m}$) in the meridional direction). While examination of the thin sections of \cite{hohn11} does suggest, in qualitative agreement with the fits, that the largest meridional stretches arise in the anterior cap, the fact that the model overestimates the actual values of these stretches may stem from the simplified modelling of the phialopore. Further, at the very latest stages of inversion (Fig.~\ref{fig:fittedshapes}j), the fitted shapes suggest very small values $f_s<0.3$ and corresponding values $f_\phi>3$ that are not borne out by the cell measurements.

\subsubsection{Phialopore Opening and Cell Rearrangement}
\begin{wrapfigure}{L}{6.3cm}
\centering\includegraphics{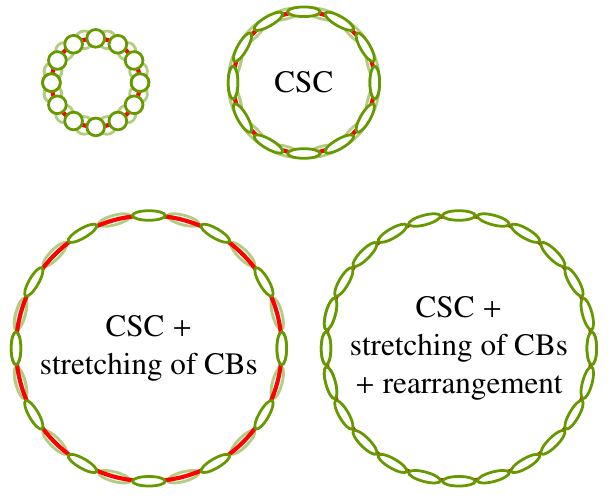}
\caption{Mechanisms of phialopore stretching: cell shape changes, stretching of cytoplasmic bridges, and cell rearrangements. Red lines represent cytoplasmic bridges; fainter colours signify other, out-of-plane cells. CSC: cell shape changes, CB: cytoplasmic bridge.}
\label{fig:phopen} 
\end{wrapfigure}

To understand how these values of the stretches at odds with the observed cell shape changes arise in the fitted shapes, we must analyse the opening of the phialopore in more detail. The observations of \cite{hohn11} show that the cytoplasmic bridges stretch considerably, to many times their initial length, as the phialopore opens. Circumferential elongation of cells as a means to increase effective radius was discussed in some detail by \cite{viamontes79}, but is not sufficient to explain the circumferential stretches observed at the phialopore. Additional elongation of cytoplasmic bridges as a means to further increase the effective radius (Fig.~\ref{fig:phopen}) may suffice to produce the large circumferential stretches, but does not explain the small values of meridional stretch at the phialopore in the fitted shapes. For this reason, we additionally imaged the opening of the phialopore using confocal laser scanning microscopy (Methods) to resolve single cells close to the phialopore (Video~2). 

\leavevmode%

The data reveal that cells rearrange near the phialopore, suggesting an additional mechanism to stretch the phialopore sufficiently for the anterior to be able to peel over the inverted posterior (Fig.~\ref{fig:phopen}). Video 2 shows how, initially, only a small number of cells form a ring at the anterior pole. When the phialopore widens, cells that were initially located away from this initial ring come to be positioned at the rim of the phialopore. It is unclear whether the cytoplasmic bridges between these cells stretch or break, or whether these cells were not connected by cytoplasmic bridges in the first place. While such cell rearrangement is beyond the scope of the current model, it is nevertheless captured qualitatively by the small values of $f_s$ near the phialopore. \cite{kelland77} observed elongation of cytoplasmic bridges near the phialopore of \emph{Volvox aureus}, but not in small fragments of broken-up embryos, and concluded that the elongation of cytoplasmic bridges was the result of passive mechanical forces. By contrast, in our model, the opening of the phialopore is the result of active cell shape changes there. This discrepancy may herald a breakdown of the approximations made to represent the phialopore.
The data also hint that there may be a different mechanical contribution at later stages of inversion (Fig.~\ref{fig:averages}i), where the rim of the phialopore may be in contact with the inverted posterior. Since the model does not resolve the rim of the phialpore in the first place, we do not pursue this further here. For completeness of the mechanical analysis, we analyse such a contact configuration in Appendix~\ref{app:1}, where we also discuss a toy problem to highlight the intricate interplay of mechanics and geometry in the contact configuration. 

\begin{figure}
\centering\includegraphics[width=0.5\textwidth]{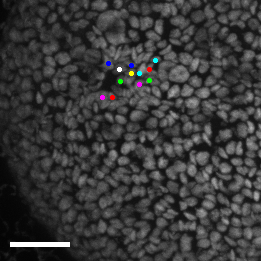}
\caption*{\textbf{Video 2.} Timelapse video of the phialopore opening obtained from confocal laser scanning microscopy of chlorophyll autofluorescence and manual tracing of selected cells (Methods). Scale bar: $20\,\text{\textmu m}$. The video shows a rearrangement of cells surrounding the phialopore.}
\label{video:phialopore}
\end{figure}

\subsection{Mechanics and Regulation of Local Shape Variations}
We now return to the spatial structure of the shape variations discussed previously. It is clear that some of this structure is geometrical: since the shapes are aligned so that the positions of their centres of mass along the axis coincide, the shape variations accumulate, and are thus expected to e.g. increase in the anterior hemisphere, towards the phialopore, as at the stage in Fig.~\ref{fig:stdshapes}c. At the same stage however, the shape variation is smaller in the bend region than in the adjacent anterior cap. Both of these regions are, however, close to the centre of mass, and so we do not expect this difference to arise from mere geometric accumulation of shape variations. We must therefore ask: can this structure arise purely mechanically (i.e. from a uniform distribution of the intrinsic parameters), but possibly as a statistical fluke, or must there be some regulation (i.e. non-uniform variation of the intrinsic parameters)?

\begin{figure}
\centering\includegraphics{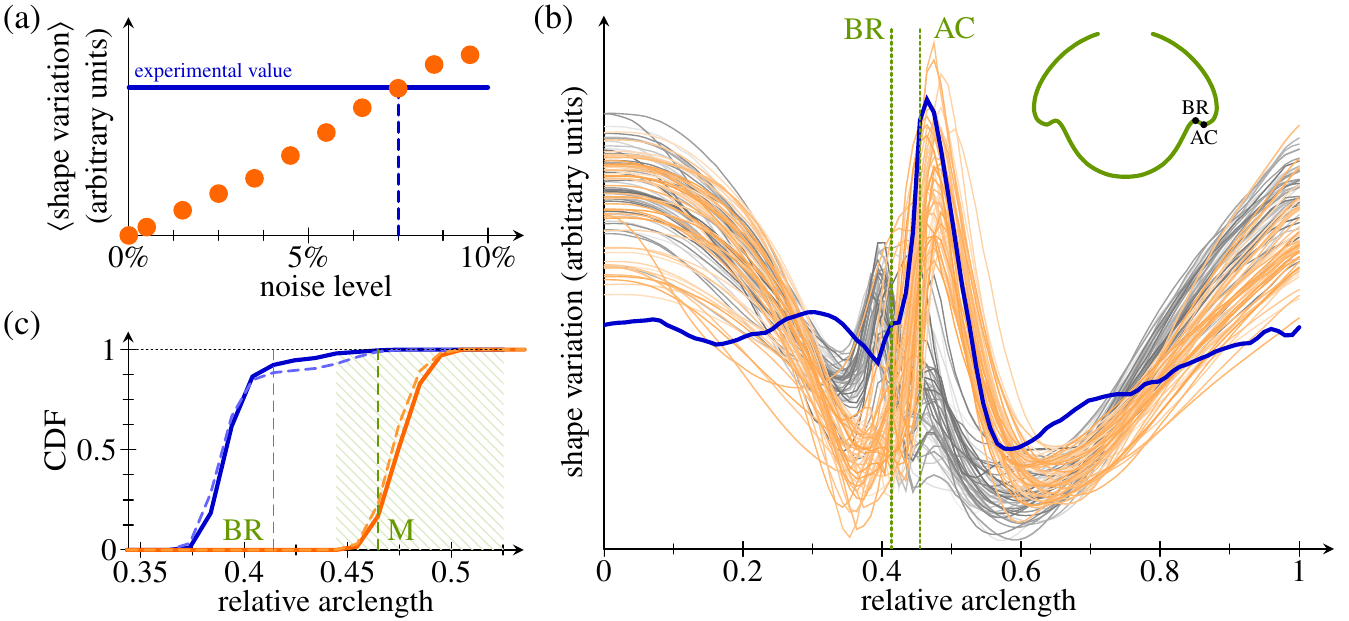}
\caption{Analysis of Shape Variations. (a)~Mean shape variation (in arbitrary units) against magnitude of uniform perturbations to the fitted shape of the stage in Fig.~\ref{fig:stdshapes}c. Each data point was obtained from 1000 perturbations of the fitted shape. Horizontal line: mean shape variation obtained from the experimental data. (b)~Magnitude of shape variations against (deformed) arclength. Thick blue line: experimental average from $N=22$ embryo halves. Thin gray lines: distributions of $N=22$ perturbations each under the uniform model. Thin orange lines: distributions of $N=22$ perturbations each under the modified model. Shape variations are scaled so that each curve has the same mean value. Inset: average shape of Fig.~\ref{fig:stdshapes}c, with bend region (BR) and anterior cap (AC) marked; these positions are marked by dotted lines in the main diagram. (c)~Cumulative distribution function (CDF) of the positions of the peak (local maximum) of shape variation under the uniform (blue lines) and modified (orange lines) models, with positions of the bend region and of the maximum (M) of experimental distribution from panel (b) labelled. Dashed lines show distributions from all random perturbations; solid lines show those from shape variations with a single local maximum. We consider a maximum to lie in the anterior cap if it falls within the hatched region, which is used for the statistical estimates in the Methods section.}
\label{fig:shapevar} 
\end{figure}

To answer this question, we analyse random perturbations of the fitted intrinsic parameters of the inversion stage in Fig.~\ref{fig:stdshapes}c. We observe that, if the relative size of perturbations (the `noise level') exceeds about $4\%$ at this stage of inversion, computation of the perturbed shapes fails for some parameter choices. This mechanical effect is not surprising: our previous analysis of invagination \citep{haas15} revealed strong shape non-linearities and the possibility of bifurcations as the magnitude of the intrinsic curvature in the bend region is increased. While we may therefore expect more leeway in some parameters than in others, we shall simply discard those perturbations for which the computation fails; further estimation of the distribution of possible perturbations is beyond the scope of the present discussion. We now estimate, for each noise level, the mean shape variation from 1000 perturbations of the fitted shape. By comparing this to the mean shape variation estimated from the $N=22$ embryo halves in Fig.~\ref{fig:averages}c, we roughly estimate a noise level of $7.5\%$ (Fig.~\ref{fig:shapevar}a). At this noise level, about $15\%$ of perturbations fail; while the non-uniformities are small, they are statistically significant (Methods).

With this noise level, we obtain 10000 samples of $N=22$ perturbations to the fitted shape each (Fig.~\ref{fig:shapevar}b), and we compute their averages in the same way as for the experimental samples. While these samples qualitatively capture the spatial structure of the shape variation, they overestimate the shape variation at the poles. More strikingly, they feature a local maximum of the shape variation in the bend region, rather than in the anterior cap. From the sample distribution of the position of these local maxima (Fig.~\ref{fig:shapevar}c), it is clear that the experimental distribution with the local maximum in the anterior cap, is very unlikely to arise under this model. We make this statement more precise statistically in the Methods section. To explain the observed structure of the shape variation, we therefore allow more variability in the meridional stretch in the anterior cap (with a noise level of $80\%$, compared to $2.5\%$ for the remaining parameters to reproduce the mean shape variation). The resulting distribution is consistent with the experimentally observed position of the local maximum of shape variation in the anterior cap (Fig.~\ref{fig:shapevar}b,c). While still overestimating the variability near the posterior pole, this modified distribution of the parameter variability captures the magnitude of the variability in the anterior cap much better than the original one. 

Thus, at this early stage of inversion (Fig.~\ref{fig:stdshapes}c), the observed embryo shapes are consistent with an increased variation of the intrinsic meridional stretch in the anterior cap. We can take the interpretation of this active regulation (or lack thereof) further by relating it to the observed cell shape changes: at the stage of Fig.~\ref{fig:stdshapes}c, the variations of the meridional stretch in the anterior cap correspond to the formation of disc-shaped cells there (Fig.~\ref{fig:volvox}d). This indicates that invagination and initiation of the expansion of the anterior hemisphere (via the formation of disc-shaped cells) are really two separate processes the relative timing of which is not crucial. (The formation of disc-shaped cells starting at different times also explains the large noise level in the meridional stretch under the modified model, although there is no fitting involved, here.) This adds to our earlier point, that these processes rely on different deformation modes (active bending for invagination and active contraction and stretching for inversion of the anterior hemisphere). These considerations also rationalise our second observation concerning the spatial structure of shape variations, that the variation in the anterior cap is reduced as inversion of the posterior hemisphere ends (Fig.~\ref{fig:stdshapes}h): there are no longer two separate processes at work. We finally point to a purely mechanical aspect of the structure of the shape variations: despite the increased variability in the anterior cap, the mechanics ensure that the variability is lowest in the bend region, where the main cell shape changes driving invagination take place.

\section{Discussion}
In this paper, we have combined experiment and theory to analyse the variability of \emph{Volvox} inversion and obtain a detailed mechanical description of this process. From observations of the structure of the variability of the shapes of inverting \emph{Volvox} embryos, we showed, using our mathematical model, that this structure results from a combination of geometry, mechanics, and active regulation. The simplest scenario with which the observed shape variations are consistent is that type-B inversion in \emph{Volvox globator} results from two separate processes, with most of the variability at the invagination stage attributed to the relative timing of these processes in individual embryos. The difference between these processes is mirrored, at a mechanical level, by the different types of deformations driving them: the first process, to invert the posterior hemisphere, mainly relies on active bending, whereas the second process, to invert the anterior hemisphere, is mainly driven by active expansion and contraction. We anticipate that these ideas and methods can be applied to other morphogenetic events in other model organisms to add to our understanding of the regulation of morphogenesis: what amount of regulation, be it spatial or temporal, of the cell-level processes is there, and how does it relate to the amount required mechanically for the processes to be able to complete? Additionally, \cite{houchmandzadeh05} showed that diffusion of two morphogens with inhibition \emph{\`a la} \cite{turing} has error-correcting properties that can explain the precise domain specification that is observed in \emph{Drosophila} embryos in spite of the huge variability of morphogen gradients \citep{houchmandzadeh02}. Does the interplay of geometry and mechanics yield analogous error-correcting properties?

While we have begun to analyse the mechanical regulation of development in the context of \emph{Volvox} inversion, our answers this far have been either negative (excluding certain mechanisms of regulation) or of what one might term the Occam's razor variety (invoking the law of parsimony to find the simplest modification of the model that can explain the observations). This approach of testing falsifiable hypotheses mitigates the risk of drawing conclusions that are mere teleology \citep{goldstein16}. Nonetheless, a fuller answer to the questions above requires estimation of the variability of the model parameters from the experimental data, yet that endeavour entails significant statistical, computational, and experimental difficulties: to estimate the variability with statistical signficance we need a large number of experimental samples to estimate the experimental distribution; for each step of the optimisation algorithm used to estimate the large number of variability parameters, a large number of computational samples must be computed to estimate the distribution under the model. Similar difficulties arise when estimating the variability allowed mechanically. While we have previously noted \citep{haas15} that the dynamic data for type-B inversion suggest that invagination proceeds without a `snap-through' bifurcation, there is no general requirement for individual developmental paths to lie on one and the same side of a mechanical bifurcation boundary. This poses an additional challenge for modelling approaches.

After this discussion of general challenges for a mechanobiological analysis of morphogenesis and its regulation, we mention some of the remaining questions specific to \emph{Volvox} inversion: our model does not resolve the details of the phialopore, and hence does not describe the closure of the phialopore at the end of inversion, which remains a combined challenge for experiment and theory: as discussed above,  the cytoplasmic bridges elongate drastically at the phialopore~\citep{hohn11}, and confocal imaging has revealed the possibility of rearrangements within the cell sheet at the phialopore. Do some cytoplasmic bridges rend to make such rearrangements possible? Understanding the details of the opening of the phialopore may also require answering a more fundamental question the answer to which has remained elusive~\citep{green81b,nishii03}: what subcellular structures are located within the cytoplasmic bridges and how is it possible for them to stretch to such an extent? At the theoretical level, rearrangements of cells near the phialopore raise more fundamental questions of morphoelasticity \citep{goriely}: in particular, how does one describe the evolution of the boundary of the manifold underlying the elastic description? Cytoplasmic bridges rending next to the phialopore would lead to the formation of lips similar to those seen in type-A inversion \citep{viamontes77,hallmann06}. Is there a simple theory to describe the elasticity of this non-axisymmetric setup?

At the close of this discussion, it is meet to briefly dwell on a question of more evolutionary flavour: how did different species of \emph{Volvox} evolve different ways of turning themselves inside out? Mapping inversion types to a phylogenetic tree of Volvocine algae shows shows that different inversion types evolved several times independently in different lineages \citep{hallmann06}. Additionally, \cite{pocock33} reported that in \emph{Volvox rousseletii} and \emph{Volvox capensis}, inversion type depends on the (sexual or asexual) reproduction mode. This may be a manifestation of the poorly understood role of environmental and evolutionary cues in morphogenesis \citep{vondassow11}, but it is natural to wonder whether there is a mechanical side to this issue. Ultimately, this is another incentive to study the mechanics of type-A inversion in more detail.  

\section{Methods and Materials}

\subsection{Acquisition of Experimental Data}
Wild-type strain \emph{Volvox globator} Linné (SAG 199.80) was obtained from the Culture Collection of Algae at the University of Göttingen, Germany~\citep{SAG}, and cultured as previously described~\citep{brumley14} with a cycle of 16\,h light at 24\textdegree C and 8\,h dark at 22\textdegree C.

\subsubsection{\textsc{OpenSPIM} Imaging}
\begin{wrapfigure}{L}{.46\textwidth}
\centering\includegraphics{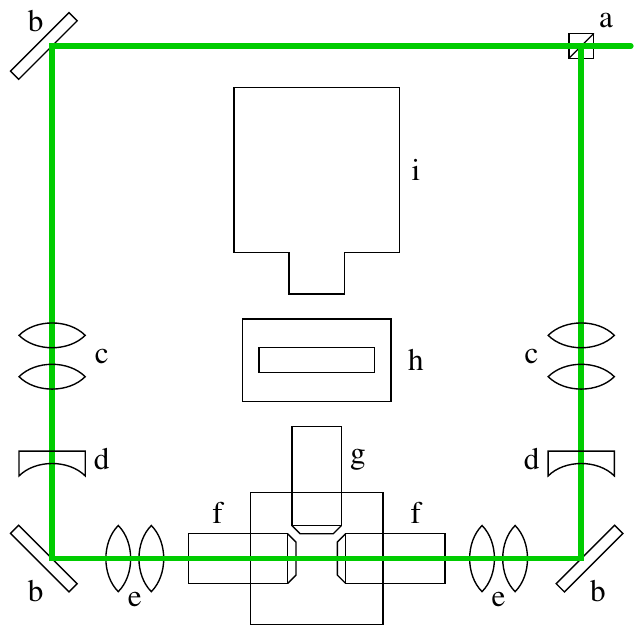}
\caption{SPIM imaging setup. a: beamsplitter cube, b: mirror, c: beam expander, d: cylindrical lens, e: telescope, f: illumination objective, g: detection objective, h: emission filter, i: camera.}
\label{fig:SPIM}
\end{wrapfigure}

A selective plane illumination microscope (SPIM) was assembled based on the \textsc{OpenSPIM} setup \citep{OpenSPIM}, with modifications to accommodate a \textsc{Stradus}\textregistered{} \textsc{Versalase}\texttrademark{} laser system with multiple wavelengths (Vortran Laser Technology, Inc., Sacramento, CA, USA) and a Cool\textsc{Snap Myo} CCD camera ($1940\times 1460$ pixels; Photometrics, AZ, USA). Moreover, to decrease the loss of data due to shadowing a second illumination arm was added to the setup (Fig.~\ref{fig:SPIM}). Illumination from both sides improved the image quality and enabled re-slicing of the z-stacks when embryos began to spin during anterior inversion. 

\emph{Volvox globator} parent spheroids were mounted in a column of low-melting-point agarose and suspended in fluid medium in the sample chamber. To visualise the cell sheet deformations of inverting \emph{Volvox globator} embryos, chlorophyll-autofluorescence was excited at $\lambda=561\,\text{nm}$ and detected at $\lambda>570\,\text{nm}$. Z-stacks were recorded at intervals of 60\,s over 4$-$6 hours to capture inversion of all embryos in a parent spheroid. We acquired time-lapse data of 13 different parent spheroids each containing 4$-$7 embryos. 

\leavevmode%

\subsubsection{Confocal Laser Scanning Microscopy}
Samples were immobilised on glass-bottom dishes by embedding them in low-melting-point agarose and covered with fluid medium. Chlorophyll-autofluorescence was excited at \mbox{$\lambda = 639\,\text{nm}$} and detected at $\lambda>647\,\text{nm}$. Z-stacks were recorded at intervals of 30\,s over 1$-$2 hours to capture inversion of a single embryo. Trajectories of individual cells close to the phialopore were obtained using \textsc{Fiji} \citep{Fiji}. Experiments were carried out using a Observer Z1 spinning-disk microscope (Zeiss, Germany).

\subsubsection{Image Tracing}
To ensure optimal image quality (traceability) for the quantitative analyses of inversion, from the inversion processes recorded with the SPIM, we selected 11 inversions (in 6 different parent spheroids) in which the acquisition plane was initially approximately parallel to the midsagittal plane of the embryos. Midsagittal cross-sections were obtained using \textsc{Fiji}~\citep{Fiji} and \textsc{Amira} (FEI, OR, USA).

Splines were fitted to these cross-sections using the following semi-automated approach implemented in Python/C++: in a preprocessing step, images were bandpass-filtered to remove short-range noise and large-range intensity correlations. Low-variance Gaussian filters were applied to smooth out the images slightly. Splines were obtained from the pre-processed images $I(\vec{x})$ using the active contour model~\citep{kass88}, with modifications to deal with intensity variations and noise in the image: the spline $\vec{x_{\text{s}}}(s)$, where $s$ is arclength, minimises an energy
\begin{equation}
\mathcal{E}[\vec{x}_{\text{s}}]=\mathcal{E}_{\text{image}}[\vec{x}_{\text{s}}]+\mathcal{E}_{\text{spline}}[\vec{x}_{\text{s}}]+\mathcal{E}_{\text{skel}}[\vec{x}_{\text{s}}], 
\end{equation}
where 
\begin{subequations}
\begin{align}
\mathcal{E}_{\text{image}}[\vec{x}_{\text{s}}] &=-\alpha\int{I\bigl(\vec{x_{\text{s}}}(s)\bigr)\,\mathrm{d}s},\\
\mathcal{E}_{\text{spline}}[\vec{x}_{\text{s}}] &= \beta\int{\biggl\|\dfrac{\partial^2\vec{x_{\text{s}}}}{\partial s^2}\biggr\|^2\mathrm{d}s}+\gamma\left(\int{\mathrm{d}s}-L_0\right)^2,\\
\mathcal{E}_{\text{skel}}[\vec{x}_{\text{s}}]&=\delta\int{I_{\text{skel}}\bigl(\vec{x_{\text{s}}}(s)\bigr)\,\mathrm{d}s},
\end{align}
\end{subequations}
wherein $\alpha,\beta,\gamma,\delta$ are parameters, $L_0$ is the estimated length of the shape outline, and $I_{\text{skel}}$ is obtained by skeletonising $I$ using the algorithm of \cite{zhang84} to minimise the number of branches. 

The energy $\mathcal{E}$ was minimised using stochastic gradient descent. Initial guesses for the splines were obtained by manually initialising about 15 timepoints for each inversion using a few guidepoints and polynomial interpolation. An initial guess for other frames was obtained from these frames by interpolation; these interpolated shapes were used to estimate $L_0$.

With $\delta = 0$, the standard active contour model of \cite{kass88} is recovered. We found that this model was not sufficient to yield fits of sufficient quality, because of the existence of local minima at small values of $\alpha$, while larger values of $\alpha$ lead to noisy splines. Thresholding methods on their own were not sufficient either, because of branching and, in particular, since they failed to capture the bend region properly.  Dynamic thresholding methods \citep[e.g.][]{otsu79} are not applicable either because of the fast variations of the brightness of the images. The modified active contour model did however produce good fits when we progressively reduced $\delta$ to zero with increasing iteration number of the minimisation scheme, yielding smooth splines, while overcoming the local minima (or, from the point of view of the skeletonisation method, choosing the correct, branchless part of the skeleton). All outlines obtained from this algorithm were manually checked and corrected.

\subsection{Analysis of Traced Embryo Shapes}
From the traced cell sheet outlines, anterior-posterior axes of the embryos were determined as follows: for shapes for which the bend region was visible on either side of the cross-section, the embryo axis was defined to be the line through the centre of mass of the shape that is perpendicular to the common tangent to the two bend regions (the apex line). Shapes were then rotated and translated manually so that their axes coincided. Since embryos do not rotate much before the flagella grow, the orientation of the axes of the earliest traces (for which the bend regions are not apparent) were taken to be the same as that of the earliest timepoint for which two bend regions were visible. The intersection of the embryo trace and axis defines the posterior pole. After manually recentring some embryos with more pronounced asymmetry, embryos were halved to obtain $N=22$ embryo halves. 

\subsubsection{Computation of Inversion Descriptors}
From the aligned shapes, the geometric descriptors of inversion reported in Fig.~\ref{fig:measurements} were computed as follows: the posterior-to-bend distance $e$ was computed as the distance from the apex line to the posterior pole. The maximal surface area $A_{\max}$ and the most negative value of curvature $\kappa_\ast$ in the bend region were computed as described previously \citep{hohn15}; traces were smoothed before computing the curvature. The phialopore width $d$ was computed as the absolute distance between the two ends of a complete embryo trace. The bend region was defined as the region of negative curvature; the distance between the first and last points of negative curvature defined the bend region width. The bend region position is defined by the distance, along the embryo trace, between the posterior pole and the midpoint of the bend region. (The latter may differ from the point where the most negative value of curvature is attained.) The values of bend region width and position obtained for each embryo half were averaged to yield the reported values.

\subsubsection{Aligning and Averaging Embryo Shapes}
To align embryos to each other, one embryo half was arbitrarily taken as the reference shape, and $T=10$ regularly spaced timepoints were chosen for fitting. (These timepoints were chosen to be well after invagination had started and before the phialopore had closed, so that defining the start and end of inversion was not required.) For each of the remaining $N-1$ embryo halves, a scale and $T$ corresponding timepoints were then sought, with shapes being (linearly) interpolated at intermediate timepoints. The interpolated and scaled shapes were centred so that the centres of mass of the cross-sections coincided. This fixes the degree of freedom of translation parallel to the embryo axis; the position perpendicular to the axis is fixed by requiring that the embryo axes coincide (Fig.~\ref{fig:averages}, figure supplement 1a). The motivation for using the centres of mass of the cross-sections (rather than of that of the embryos, which assigns the same mass to each cell by assigning more mass to those points of the cross-section that are farther away from the embryo axis) is a biological one: because of the cylindrical symmetry of the cell shape changes, this average assigns the same mass to each cell shape change. 

For aligning embryo shapes, we distribute $M=100$ averaging points uniformly along the (possibly different) arclength of each of the embryo halves. Corresponding points were determined using dynamic time warping (DTW) as described by e.g. \cite{dtw}, and the distances between these shapes and their averages were minimised as explained in what follows. The parameters describing the alignment are thus the scale factors $S_1 = 1, S_2,\dots S_N$ and the averaging time points $\vec{\tau}_1 = (\tau_{11},\tau_{12},\dots,\tau_{1T}),\vec{\tau}_2,\dots,\vec{\tau}_N$, where $\vec{\tau}_1$ is fixed. Each choice of these parameters yields a set of shapes $\vec{X}_1=(x_{11},\dots,x_{1M}),\vec{X}_2,\dots,\vec{X}_N$ with points matched up by maps $\sigma_1,\sigma_2,\dots,\sigma_N$ obtained from the DTW algorithm. The effect of the local stretching allowed by the DTW algorithm is illustrated in Fig.~\ref{fig:averages}, figure supplement 1b,c. The mean shapes having been determined, the sum of Euclidean distances between shapes of individual embryos and the mean,
\begin{equation}
\sum_{t=1}^T{\left\{\sum_{n=1}^N{\sum_{m=1}^M}{\left(x_{n\sigma_n(m)}-\overline{x}_m\right)^2}\right\}^{1/2}},\qquad\mbox{where }\overline{x}_m=\dfrac{1}{N}\sum_{n=1}^N{x_{n\sigma_n(m)}},\label{eq:score} 
\end{equation}
was minimised over the space of all these alignment parameters using the \textsc{Matlab} (The MathWorks, Inc.) routine \texttt{fminsearch}, modified to incorporate the variant of the Nelder--Mead algorithm suggested by \cite{gao12} for problems with a large number of parameters. After the algorithm had converged, each of the alignment parameters was modified randomly, and the algorithm was run again. This was repeated until the alignment score defined by (\ref{eq:score}) did not decrease further. The means $\overline{x}_1,\overline{x}_2,\dots,\overline{x}_M$ for the alignment minimising (\ref{eq:score}) define the average embryo shapes.

Aligning shapes in this way using dynamic time warping requires a considerable amount of computer time. To make the problem computationally tractable, we invoked the usual heuristics of only computing pairwise DTW distances, and reducing the size of the DTW matrix by only computing a band centred on the diagonal. To verify the algorithm, we also ran several instantiations of the alignment algorithm without DTW (i.e. with $\sigma_n=\mathrm{id}$) and with larger parameter randomisations, confirming that the modified Nelder--Mead algorithm finds an appropriate alignment. This also enabled us to verify that results do not change qualitatively if the centres of mass of the cross-sections are replaced with those of the embryo halves (even though, as noted in the main text, the shapes without DTW are unsatisfactory since they have kinks in the bend region that are not seen in individual embryo shapes).

For the simple alternative averaging method in Fig.~\ref{fig:averages}, figure supplement 2, different numbers of averaging points were distributed at equal arclength spacing along all individual shapes. Differences in arclengths of individual embryos mean that the rims of the phialopores of individual embryo halves are not necessarily matched up (Fig.~\ref{fig:averages}, figure supplement 1c). No time stretching was applied. The averaging method in Fig.~\ref{fig:averages}, figure supplement 3, is the method discussed above, without DTW (i.e. with $\sigma_n=\mathrm{id}$).

\subsection{Elastic Model}

\begin{figure}
\centering\includegraphics{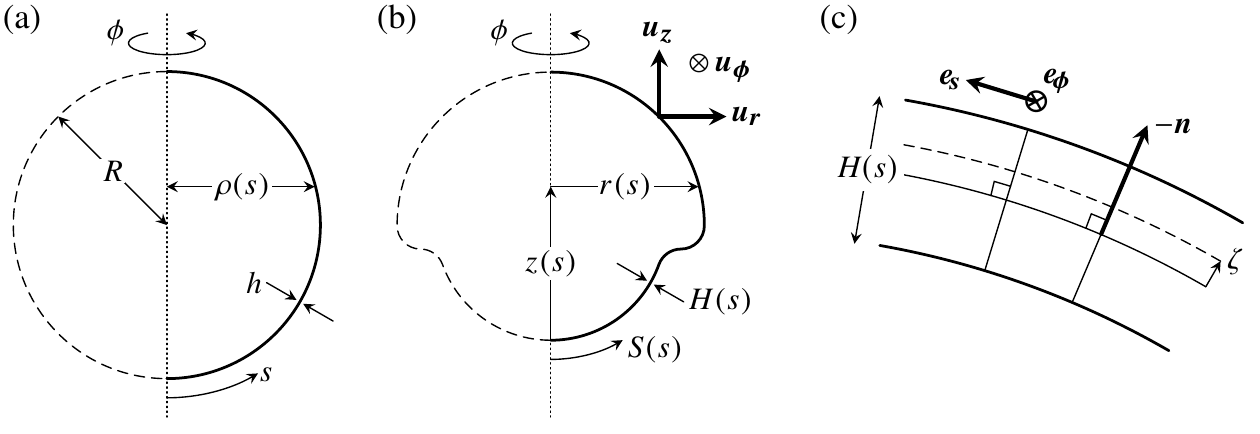}
\caption{Geometry of the problem. (a)~Undeformed geometry: a spherical shell of radius $R$ and thickness $h\ll R$ is characterised by its arclength $s$ and distance from the axis of revolution $\rho(s)$. (b)~Deformed configuration, characterised by its arclength $S(s)$ and distance $r(s)$ from the axis of revolution. Intrinsic volume conservation sets the thickness $H(s)$ of the sheet. A local basis $(\vec{u_r},\vec{u_\phi},\vec{u_z})$ describes the deformed surface. (c)~Cross-section of the shell under the Kirchhoff hypothesis, with a coordinate $\zeta$ across the thickness of the shell, parallel to the normal $\vec{n}$ to the midsurface.}
\label{figS1}
\end{figure}

We consider a spherical shell of radius $R$ and uniform thickness $h\ll R$ (Fig.~\ref{figS1}a), characterised by its arclength $s$ and distance from the axis of revolution $\rho(s)$, to which correspond arclength $S(s)$ and distance from the axis of revolution $r(s)$ in the axisymmetric deformed configuration (Fig.~\ref{figS1}b). We define the meridional and circumferential stretches
\begin{align}
&f_s(s)=\dfrac{\mathrm{d}S}{\mathrm{d}s},&& f_\phi(s)=\dfrac{r(s)}{\rho(s)}.
\end{align}
The position vector of a point on the midsurface of the deformed shell is thus
\begin{equation}
\vec{r}(s,\phi)=r(s)\vec{u_r}(\phi)+z(s)\vec{u_z}, 
\end{equation}
in a right-handed set of axes $(\vec{u_r},\vec{u_\phi},\vec{u_z})$ and so the tangent vectors to the deformed midsurface are
\begin{align}
\vec{e_s}&=r'\vec{u_r}+z'\vec{u_z},&&\vec{e_\phi}=r\vec{u_\phi},
\end{align}
where dashes denote differentiation with respect to $s$. By definition, $r'^2+z'^2=f_s^2$, and so we may write 
\begin{align}
&r'=f_s\cos{\beta},&& z'=f_s\sin{\beta}.
\end{align}
Hence the normal to the deformed midsurface is
\begin{equation}
\vec{n}= \dfrac{r'\vec{u_z}-z'\vec{u_r}}{f_s}=\cos{\beta}\,\vec{u_z}-\sin{\beta}\,\vec{u_r}.
\end{equation}
We now make the Kirchhoff `hypothesis'~\citep{audolypomeau}, that the normals to the undeformed midsurface remain normal to the deformed midsurface (Fig.~\ref{figS1}c). Taking a coordinate $\zeta$ across the thickness $h$ of the undeformed shell, the position vector of a general point in the shell is
\begin{equation}
\vec{r}(s,\phi,\zeta)=r\vec{u_r}+z\vec{u_z}+\zeta\vec{n}=(r-\zeta\sin{\beta})\vec{u_r}+(z+\zeta\cos{\beta})\vec{u_z}.\label{eq:ds}
\end{equation}
The tangent vectors to the shell are thus
\begin{align}
\vec{e_s}&= f_s(1-\kappa_s\zeta)(\cos{\beta}\,\vec{u_r}+\sin{\beta}\,\vec{u_z}),&\vec{e_\phi}&= \rho f_\phi(1-\kappa_\phi\zeta)\vec{u_\phi},
\end{align}
where $\kappa_s=\beta'/f_s$ and $\kappa_\phi=\sin{\beta}/r$ are the curvatures of the deformed midsurface. The metric of the deformed shell under the Kirchhoff hypothesis accordingly takes the form
\begin{equation}
\mathrm{d}\boldsymbol{r}^2 = f_s^2(1-\kappa_s\zeta)^2\mathrm{d}s^2+f_\phi^2(1-\kappa_\phi\zeta)^2\rho^2\mathrm{d}\phi^2.
\end{equation}
The geometric and intrinsic deformation gradient tensors are thus
\begin{align}
&\mathbfsfit{F}^{\mathbfsf{g}}=\left(\begin{array}{cc}
f_s(1-\kappa_s\zeta)&0\\
0&f_\phi(1-\kappa_\phi\zeta)
\end{array}\right), &&\mathbfsfit{F^0}=\left(\begin{array}{cc}
f_s^0(1-\kappa_s^0\zeta)&0\\
0&f_\phi^0(1-\kappa_\phi^0\zeta)
\end{array}\right),\label{eq:defgrad}
\end{align}
where $f_s^0,f_\phi^0$ and $\kappa_s^0,\kappa_\phi^0$ are the intrinsic stretches and curvatures of the shell. Thence, invoking the standard multiplicative decomposition of morphoelasticity~\citep{goriely}, the elastic deformation gradient tensor is
\begin{equation}
\mathbfsfit{F}=\mathbfsfit{F}^{\mathbfsf{g}}\bigl(\mathbfsfit{F^0}\bigr)^{-1}=\left(\begin{array}{cc}
\dfrac{f_s(1-\kappa_s\zeta)}{f_s^0(1-\kappa_s^0\zeta)}&0\\
0&\dfrac{f_\phi(1-\kappa_\phi\zeta)}{f_\phi^0(1-\kappa_\phi^0\zeta)}
\end{array}\right).
\end{equation}
While we do not make any assumption about the \emph{geometric} or \emph{intrinsic} strains derived from $\mathbfsfit{F}^{\mathbfsf{g}}$ and $\mathbfsfit{F^0}$, respectively, we assume that the \emph{elastic} strains derived from $\mathbfsfit{F}$ remain small; we may thus approximate 
\begin{align}
&\varepsilon_{ss}\approx \dfrac{f_s(1-\kappa_s\zeta)}{f_s^0(1-\kappa_s^0\zeta)}-1, &&\varepsilon_{\phi\phi}\approx \dfrac{f_s(1-\kappa_s\zeta)}{f_s^0(1-\kappa_s^0\zeta)}-1,
\end{align}
with the off-diagonal elements vanishing, $\varepsilon_{s\phi}=\varepsilon_{\phi s}=0$. For a Hookean material with elastic modulus $E$ and Poisson's ratio~$\nu$ \citep{libai,audolypomeau}, the elastic energy density (per unit extent in the meridional direction) is found by integrating across the thickness of the shell:
\begin{align}
\dfrac{\mathcal{E}}{2\pi\rho}&=\dfrac{E}{2(1-\nu^2)}\int_{-h/2}^{h/2}{\Bigl(\varepsilon_{ss}^2+\varepsilon_{\phi\phi}^2+2\nu\varepsilon_{ss}\varepsilon_{\phi\phi}\Bigr)\mathrm{d}\zeta}\nonumber\\
&=\dfrac{Eh}{2(1-\nu^2)}\biggl\{\left(1+\dfrac{h^2}{4}{\kappa_s^0}^2\right)E_s^2+\left(1+\dfrac{h^2}{4}{\kappa_\phi^0}^2\right)E_\phi^2+2\nu\left(1+\dfrac{h^2}{12}\left({\kappa_s^0}^2+\kappa_s^0\kappa_\phi^0+{\kappa_\phi^0}^2\right)\right)E_sE_\phi\biggr\}\nonumber\\
&\qquad+\dfrac{Eh^3}{24(1-\nu^2)}\biggl\{K_s^2+K_\phi^2+2\nu K_sK_\phi-4\kappa_s^0E_sK_s-4\kappa_\phi^0E_\phi K_\phi-2\nu\bigl(\kappa_s^0+\kappa_\phi^0\bigr)\bigl(E_\phi K_s+E_sK_\phi\bigr)\biggr\},\label{eq:E}
\end{align}
where we have expanded the energy up to third order in the thickness, and where we have defined the shell strains and curvature strains
\begin{align}
&E_s=\dfrac{f_s-f_s^0}{f_s^0},&&E_\phi=\dfrac{f_\phi-f_\phi^0}{f_\phi^0},  &&K_s=\dfrac{f_s\kappa_s-f_s^0\kappa_s^0}{f_s^0},&&K_\phi=\dfrac{f_\phi\kappa_\phi-f_\phi^0\kappa_\phi^0}{f_\phi^0}.   
\end{align}
As in our previous work~\citep{hohn15,haas15}, the elastic modulus is an overall constant that ensures that $\mathcal{E}$ has units of energy, but does not otherwise affect the shapes. We shall also assume that $\nu=1/2$ for incompressible biological material; the cell size measurements of \cite{viamontes79} for type-A inversion in \emph{Volvox carteri} support this assumption qualitatively. (These considerations also explain why we do not perturb these mechanical parameters in our analysis of the shape variations.) We finally set $h/R=0.15$ as in our previous work.

\subsubsection{Derivation of the Governing Equations}
The derivation of the governing equations proceeds similarly to standard shell theories~\citep{libai,audolypomeau,knoche11}. In fact, the resulting equations turn out to have a form very similar to those of standard shell theories, but a host of extra terms arise in the expressions for the shell stresses and moments due to the assumptions of morphoelasticity. The variation of the elastic energy takes the form
\begin{equation}
\dfrac{\delta\mathcal{E}}{2\pi\rho}=n_s\delta E_s+n_\phi\delta E_\phi+m_s\delta K_\phi+m_\phi\delta K_\phi,
\end{equation}
with\begin{subequations}
\begin{align}
\delta E_s&=\dfrac{\delta f_s}{f_s^0}=\dfrac{1}{f_s^0}\Bigl(\sec{\beta}\,\delta r'+f_s\tan{\beta}\,\delta\beta\Bigr),&\delta E_\phi&=\dfrac{\delta f_\phi}{f_\phi^0}=\dfrac{\delta r}{f_\phi^0\rho},\\
\delta K_s&=\dfrac{\delta(f_s\kappa_s)}{f_s^0}=\dfrac{\delta\beta'}{f_s^0},&\delta K_\phi&=\dfrac{\delta(f_\phi\kappa_\phi)}{f_\phi^0}=\dfrac{\cos{\beta}}{f_\phi^0\rho}\delta\beta,
\end{align}
\end{subequations}
wherein dashes again denote differentiation with respect to $s$, and where the shell stresses and moments are defined by\begin{subequations}
\begin{align}
n_s&=\dfrac{Eh}{1-\nu^2}\left\{E_s+\nu E_\phi+\dfrac{h^2}{12}\left(3{\kappa_s^0}^2E_s+\nu\left({\kappa_s^0}^2+\kappa_s^0\kappa_\phi^0+{\kappa_\phi^0}^2\right)E_\phi-2\kappa_s^0K_s-\nu\bigl(\kappa_s^0+\kappa_\phi^0\bigr)K_\phi\right)\right\},\label{eq:ns}\\
n_\phi&=\dfrac{Eh}{1-\nu^2}\left\{E_\phi+\nu E_s+\dfrac{h^2}{12}\left(3{\kappa_\phi^0}^2E_\phi+\nu\left({\kappa_s^0}^2+\kappa_s^0\kappa_\phi^0+{\kappa_\phi^0}^2\right) E_s-2\kappa_\phi^0K_\phi-\nu\bigl(\kappa_s^0+\kappa_\phi^0\bigr)K_s\right)\right\},
\end{align}
\end{subequations}
and
\begin{subequations}
\begin{align}
m_s&=\dfrac{Eh^3}{12(1-\nu^2)}\biggl\{K_s+\nu K_\phi-2\kappa_s^0 E_s-\nu\bigl(\kappa_s^0+\kappa_\phi^0\bigr)E_\phi\biggr\},\label{eq:ms}\\
m_\phi&=\dfrac{Eh^3}{12(1-\nu^2)}\biggl\{K_\phi+\nu K_s-2\kappa_\phi^0 E_\phi-\nu\bigl(\kappa_s^0+\kappa_\phi^0\bigr)E_s\biggr\}.
\end{align}
\end{subequations}
Defining
\begin{align}
&N_s=\dfrac{n_s}{f_s^0f_\phi},&& N_\phi=\dfrac{n_\phi}{f_\phi^0f_s},&&M_s=\dfrac{m_s}{f_s^0f_\phi},&& M_\phi=\dfrac{m_\phi}{f_\phi^0f_s},
\end{align}
the variation becomes
\begin{align}
\dfrac{\delta\mathcal{E}}{2\pi}&=\Bigl\llbracket r N_s\sec{\beta}\,\delta r+r M_s\,\delta\beta\Bigr\rrbracket\nonumber\\
&\hspace{10mm}-\int{\biggl\{\biggl(\dfrac{\mathrm{d}}{\mathrm{d}s}\Bigl(r N_s\sec{\beta}\Bigr)-f_sN_\phi\biggr)\delta r - \biggl(r f_sN_s\tan{\beta}}+f_sM_\phi\cos{\beta}-\dfrac{\mathrm{d}}{\mathrm{d}s}\Bigl(r M_s\Bigr)\biggr)\delta\beta\biggr\}\,\mathrm{d}s. \label{eq:var}
\end{align}
The Euler--Lagrange equations of (\ref{eq:E}) are thus
\begin{align}
&\dfrac{\mathrm{d}}{\mathrm{d}s}\Bigl(rN_s\sec{\beta}\Bigr)-f_sN_\phi=0, &&\dfrac{\mathrm{d}}{\mathrm{d}s}\Bigl(rM_s\Bigr)-f_sM_\phi\cos{\beta}-rf_sN_s\tan{\beta}=0.\label{eq:E2}
\end{align}
To remove the singularity that arises in the second of (\ref{eq:E2}) when $\beta=\pi/2$, we define the transverse shear tension $T=-N_s\tan{\beta}$ as in standard shell theories. The governing equations can then be rearranged to give
\begin{align}
\dfrac{\mathrm{d}N_s}{\mathrm{d}s}&=f_s\left(\dfrac{N_\phi-N_s}{r}\cos{\beta}+\kappa_s T\right),&\dfrac{\mathrm{d}M_s}{\mathrm{d}s}&=f_s\left(\dfrac{M_\phi-M_s}{r}\cos{\beta}-T\right).\label{eq:Eb}
\end{align}
By differentiating the definition of $T$ and using the first of (\ref{eq:Eb}), one finds that
\begin{equation}
\dfrac{\mathrm{d}T}{\mathrm{d}s}=-f_s\left(\kappa_sN_s+\kappa_\phi N_\phi+\dfrac{T}{r}\cos{\beta}\right). \label{eq:ET}
\end{equation}
Together with the geometrical equations $r'=f_s\cos{\beta}$ and $\beta'=f_s\kappa_s$, equations (\ref{eq:Eb}) and (\ref{eq:ET}) describe the deformed shell. The five required boundary conditions can be read off the variation (\ref{eq:var}) and the definition of $T$,
\begin{subequations}
\begin{align}
\beta &= 0, &r&= 0, &&T = 0&&\text{at the posterior pole},\\
N_s &= 0, &M_s &= 0&&&&\text{at the phialopore}.
\end{align}
\end{subequations}
We solve these equations numerically using the boundary value-problem solver \texttt{bvp4c} of \textsc{Matlab} (The MathWorks, Inc.). 

For completeness, we note that if external forces are applied to the shell, and $\delta\mathcal{W}$ is the variation of the work done by these forces, then the variational condition is $\delta\mathcal{E}+\delta\mathcal{W}=0$. In that case, it is useful to write the variation (\ref{eq:var}) in terms of $\delta r$ and $\delta z$. We note that $\delta r'=-f_s\sin{\beta}\,\delta \beta$ and $\delta z'=f_s\cos{\beta}\,\delta \beta$, and so
\begin{equation}
f_s\delta\beta=\cos{\beta}\,\delta z'-\sin{\beta}\,\delta r'.
\end{equation}
Using this geometric relation and integrating by parts, we obtain
\begin{align}
\dfrac{\delta\mathcal{E}}{2\pi}&=\Biggl\llbracket r M_s\,\delta\beta+\left\{rN_s\cos{\beta}-\dfrac{\sin{\beta}}{f_s}\left(M_\phi\cos{\beta}-\dfrac{\mathrm{d}}{\mathrm{d}s}\Bigl(rM_s\Bigr)\right)\right\}\delta r\nonumber\\
&\hspace{45mm}+\left\{rN_s\sin{\beta}+\dfrac{\cos{\beta}}{f_s}\left(M_\phi\cos{\beta}-\dfrac{\mathrm{d}}{\mathrm{d}s}\Bigl(rM_s\Bigr)\right)\right\}\delta z\Biggr\rrbracket\nonumber\\
&\hspace{15mm}+\int{\Biggl\{f_sN_\phi-\dfrac{\mathrm{d}}{\mathrm{d}s}\left(rN_s\cos{\beta}-\dfrac{\sin{\beta}}{f_s}\left(M_\phi\cos{\beta}-\dfrac{\mathrm{d}}{\mathrm{d}s}\Bigl(rM_s\Bigr)\right)\right)\Biggr\}\,\delta r\,\mathrm{d}s}\nonumber \\
&\hspace{30mm}-\int{\dfrac{\mathrm{d}}{\mathrm{d}s}\left(rN_s\sin{\beta}+\dfrac{\cos{\beta}}{f_s}\left\{M_\phi\cos{\beta}-\dfrac{\mathrm{d}}{\mathrm{d}s}\Bigl(rM_s\Bigr)\right\}\right)\delta z\,\mathrm{d}s}.\label{eq:var2}
\end{align}

\subsubsection{Limitations of the Theory}
The theory presented here has a singularity in a biologically relevant limit: the intrinsic deformation gradient $\mathbfsfit{F^0}$ becomes singular at $|\kappa_s^0|=(h/2)^{-1}$ or $|\kappa_\phi^0|=(h/2)^{-1}$. This value corresponds precisely to the case of cells that are constricted to a point at one cell pole. 

The way around this issue would presumably involve writing down an energy directly relative to the (possibly incompatible) intrinsic configuration of the shell. Working in the intrinsic configuration of the shell raises another issue to contend with, however: intrinsic volume conservation, which implies that the thickness $H$ of the intrinsically deformed shell, which is close to the thickness of the deformed shell by assumption, differs from the thickness $h$ of the undeformed shell. For a doubly curved shell, the relative thickness $\eta=H/h$ is a function of both the intrinsic stretches $f_s^0,f_\phi^0$ and the intrinsic curvatures $\kappa_s^0,\kappa_\phi^0$. The volume of an element of shell is
\begin{equation}
\int_{-H/2}^{H/2}{f_s^0f_\phi^0\bigl(1-\kappa_s^0\zeta\bigr)\bigl(1-\kappa_\phi^0\zeta\bigr)\rho\,\mathrm{d}s\,\mathrm{d}\phi\,\mathrm{d}\zeta}= f_s^0f_\phi^0H\left(1+\dfrac{H^2}{12}\kappa_s^0\kappa_\phi^0\right)\rho\,\mathrm{d}s\,\mathrm{d}\phi. 
\end{equation}
It follows that $\eta$ satisfies satisfies the cubic equation
\begin{equation}
\left(\dfrac{h^2}{12}f_s^0f_\phi^0\kappa_s^0\kappa_\phi^0\right)\eta^3+f_s^0f_\phi^0\eta-\left(1+\dfrac{h^2}{12R^2}\right)=0, \label{eq:eta}
\end{equation}
the solution of which can be expressed in closed form. It is clear that this equation always has a solution if $\kappa_s^0\kappa_\phi^0>0$. If $\kappa_s^0\kappa_\phi^0<0$, there is a solution if and only if
\begin{equation}
\bigl|\kappa_s^0\kappa_\phi^0\bigr|<\left(\dfrac{4f_s^0f_\phi^0}{3h}\right)^2\left(1+\dfrac{h^2}{12R^2}\right)^{-2}. 
\end{equation}
Since $16/9<4$, this condition may fail before the intrinsic geometry becomes singular, so this additional condition is not vacuous. This brief discussion therefore points to some interesting, more fundamental problems in the theory of morphoelastic shells.

There is an additional subtlety associated with the geometric and intrinsic deformation gradient tensors in Eq.~(\ref{eq:defgrad}): the components of $\mathbfsfit{F}^{\mathbfsf{g}}$ are expressed in (\ref{eq:defgrad}) relative to the (natural) mixed basis $\bigl\{\vec{\hat{e}_s},\vec{\hat{e}_\phi}\bigr\}\otimes\bigl\{\vec{\hat{E}_s},\vec{\hat{E}_\phi}\bigr\}$, where $\vec{\hat{e}_s},\vec{\hat{e}_\phi}$ are the unit vectors tangent to the deformed configuration of the shell and $\vec{\hat{E}_s},\vec{\hat{E}_\phi}$ are defined analogously for the undeformed configuration. We have implicitly written down the components of $\mathbfsfit{F^0}$ relative to the same basis. In general however, the components of $\mathbfsfit{F^0}$ in (\ref{eq:defgrad}) are those relative to the basis $\bigl\{\vec{\hat{e}^0_s},\vec{\hat{e}^0_\phi}\bigr\}\otimes\bigl\{\vec{\hat{E}_s},\vec{\hat{E}_\phi}\bigr\}$, where the unit basis $\bigl\{\vec{\hat{e}^0_s},\vec{\hat{e}^0_\phi}\bigr\}$ can \emph{a priori} be specified freely. We have neglected these additional degrees of freedom in the above derivation; the question of how to define a natural intrinsic tangent basis $\bigl\{\vec{\hat{e}^0_s},\vec{\hat{e}^0_\phi}\bigr\}$ is however an interesting one, since the intrinsic stretches and curvatures need not be compatible.

\subsection{Fitting Embryo Shapes}
\begin{figure}
\centering\includegraphics{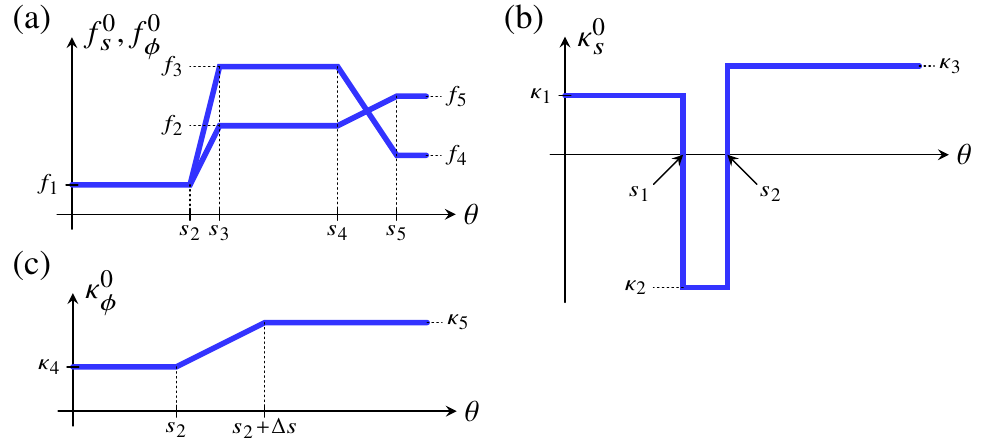}
\caption{Shape Fitting. Piecewise constant or linear functional forms of (a) the intrinsic stretches $\smash{f_s^0,f_\phi^0}$, (b) the meridional intrinsic curvature $\kappa_s^0$, and (c) the circumferential intrinsic curvature $\smash{\kappa_\phi^0}$, plotted against the arclength $s$ of the undeformed shell. Labels define fifteen fitting parameters. The constant $\Delta s = 0.05$ is set arbitrarily for continuity.}
\label{figSfit}
\end{figure}

For the purpose of fitting the model to the observed averages shapes, we define a family of piecewise constant or  linear functional forms for the intrinsic stretches and curvatures, shown in Fig.~\ref{figSfit}. This family of intrinsic stretches and curvatures is defined in terms of fifteen parameters, which are to be fitted for. Their functional forms are based on observations of cell shape changes by \cite{hohn11} summarised below: 
\begin{itemize}[leftmargin=*]
\item The intrinsic stretches $f_s^0,f_\phi^0$ vary in the both hemispheres (Fig.~\ref{figSfit}a): in the posterior hemisphere, the initially teardrop-shaped cells thin into spindle-shaped cells (Fig.~\ref{fig:volvox}c,d, Fig.~\ref{fig:cellshapechanges}b), while, in the anterior hemisphere, they flatten into disc-shaped (`pancake-shaped') cells (Fig.~\ref{fig:volvox}d,e, Fig.~\ref{fig:cellshapechanges}c). While the evolution towards spindle-shaped cells appears to occur at the same time all over the posterior hemisphere, the data from thin sections suggest that the transition to disc-shaped cells starts at the bend region and progresses towards the phialopore (Fig.~\ref{fig:volvox}d,e). Moreover, the spindle-shaped cells are isotropic, $f_s^0\approx f_\phi^0$, while the pancake-shaped cells are markedly anisotropic: next to the bend region, the long axis of their elliptical cross-section is the meridional one; next to the phialopore, it is the circumferential axis (Fig.~\ref{fig:cellshapechanges}c). 
\item The meridional intrinsic curvature $\kappa_s^0$ (Fig.~\ref{figSfit}b) is expected to vary most drastically in the region where paddle-shaped cells with thin wedge ends form (Fig.~\ref{fig:volvox}d, Fig.~\ref{fig:cellshapechanges}a). Because of the motion of cytoplasmic bridges relative to the cells, some additional, yet slighter, variation may be expected.
\item The variations of the circumferential intrinsic curvature $\kappa_\phi^0$ are less clear: on the one hand, $\kappa_\phi^0$ does not vary as drastically as the meridional one, because of the anisotropy of the paddle-shaped cells. This is a marked difference to type-A inversion, where the flasks-shaped cells are isotropic \citep{viamontes79}, and both intrinsic curvatures therefore vary more dramatically in the bend region. On the other hand, some variation of the circumferential intrinsic curvature may be expected because of the motion of cytoplasmic bridges (Fig.~\ref{figSfit}c). We impose a continuous functional form for $\kappa_\phi^0$, regularising a step function over a distance $\Delta s$ in arclength (Fig.~\ref{figSfit}c), but we do not fit for $\Delta s$ since we lack detailed information about the cell shape changes that define it.
\end{itemize}
The other geometrical parameter of the shell, the angular extent $P$ of the phialopore, is not fitted for. We arbitrarily set $P=0.3$. The reasons for this simplification are discussed in the main text.

Numerical shapes were fitted to the average shapes by distributing $M=100$ points uniformly along the arclength of the numerical and average shapes, and minimising a Euclidean distance between them using \textsc{Matlab} (The MathWorks, Inc.) routine \texttt{fminsearch}, modified as discussed above. A custom-written adaptive stepper was used to move about in parameter space and select the initial guess for the Nelder--Mead simplex. For each shape, the fit for the previous stage of inversion was used as the initial guess for the optimisation.

\subsection{Shape Perturbations and Statistical Statements}
To define perturbations for the $F=15$ fitted model parameters $\vec{P_0}\in\mathbb{R}^F$ at noise level $\delta$, we draw independent $N$ uniform random samples $\vec{X}\sim\mathcal{U}[0,1]^F$ on the unit interval and define the perturbed parameters $\vec{P}=\vec{P_0}\bigl(1+2\delta(\vec{X}-1)\bigr)$.
\subsubsection{Uniformity of the Distribution of Perturbations}
As discussed in the main text, some of these perturbed parameters must be discarded. As a result, the samples that are retained are uniform on an unknown set $\mathcal{A}\subseteq[0,1]^F$ with means~$\vec{\mu}$. To establish that these means are not all the same, we derive confidence intervals for $\mu_i-\mu_j$. Since $|X_i-X_j|\leqslant 1$, we may bound the variance of these differences by $\mathrm{Var}(X_i-X_j)\leqslant 1$, and hence, by the central limit theorem, a $100(1-p)\%$ confidence interval is
\begin{equation}
\langle X_i\rangle-\langle X_j\rangle\pm\dfrac{z}{\sqrt{N}},\quad\mbox{where }z=\Phi^{-1}\left(1-\dfrac{p/2}{\binom{F}{2}}\right), 
\end{equation}
wherein $\Phi^{-1}$ is the inverse of the cumulative distribution function of the $\mathcal{N}(0,1)$ distribution, and where have included a multiple-testing correction. At noise level $\delta=0.075$, we have run 10000 perturbations, finding $M=\max{\langle\vec{X}\rangle}\approx 0.526$ and $m=\min{\langle\vec{X}\rangle}\approx 0.485$. With $M-m\approx 0.041$ and \mbox{$\Phi^{-1}(1-0.005/105)/\sqrt{N}\approx 0.039$}, we infer that the 99\% confidence interval for the maximum difference of the means does not contain zero, and hence that the means are not all the same. We notice however that these deviations of the means are small, in that they are not statistically signficantly different from $0.5$. 
\subsubsection{Position of the Maxima of Shape Variation}
We now make quantitative our statement, based on the cumulative distributions in Fig.~\ref{fig:shapevar}c, that the experimental distribution of shape variation (with a maximum in the anterior cap) is very unlikely to arise under the uniform model. We ask: what is the probability $p$, under the uniform model, for the maximum in shape variation to lie in the anterior cap (Fig.~\ref{fig:shapevar}c)? For 10000 perturbations, we found that 757 had a maximum in the anterior cap. Among these perturbations, 2345 yielded a single maximum in shape variation, with 60 of these maxima in the anterior cap. With 99\% confidence, we therefore have upper bounds $p<0.0757+0.0129<0.09$ from all perturbations, and $p<0.0256+0.0266<0.06$ if we restrict to shape variations with a single maximum. 

\section{Acknowledgements}
We are grateful to D. Page-Croft and C. Hitch for instrument fabrication. We thank S. Hilgenfeldt for asking a question at the right time, T. B. Berrett for a conversation on matters statistical, and the Engineering and Physical Sciences Research Council, the Schlumberger Chair Fund, and the Wellcome Trust for partially funding this work.

%\nocite{*} % This command displays all refs in the bib file
\bibliography{inv2}

\appendix
\begin{appendixbox}\label{app:1}
In this appendix, we analyse the configuration where the rim of the phialopore is in contact with the inverted posterior for completeness of the mechanical analysis. We also analyse a toy problem to illustrate the intricate interplay of geometry and mechanics during contact.
\renewcommand{\theequation}{A\arabic{equation}}
\setcounter{equation}{0}
\subsection{Elastic Model in the Contact Configuration}
Let $P$ be the angular extent of the axisymmetric phialopore at the anterior pole of the shell. Here, we discuss the contact problem where the shell has deformed in such a way that the rim of the phialopore (at $\theta=\pi-P=Q$, where $\theta=s/R$ is the polar angle) is in contact with the shell at some as yet unknown position $\theta=C$, as shown in Fig.~\ref{figS2}a,b. 

\begin{center}
\includegraphics{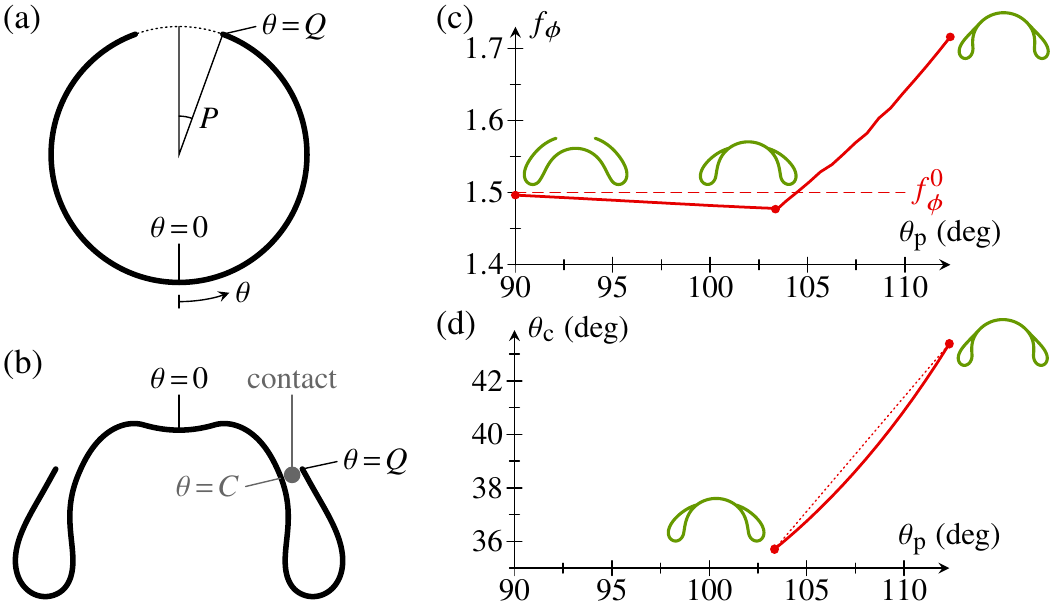}
\captionof{figure}{Analysis of the contact problem. (a) Undeformed configuration and (b) contact configuration. The phialopore at $\theta=Q=\pi-P$ touches the shell at $\theta=C$, where $\theta$ is the polar angle. (c)~Increasing circumferential stretch $\smash{f_\phi}$ with advancing position $\theta_{\mathrm{p}}$ of the peeling front, at constant intrinsic stretch $\smash{f_\phi^0}$. Insets: configuration with inverted posterior (as in Fig.~\ref{fig:antpeeling}d), at beginning of contact, and at a later stage. (d)~Advancing contact position with advancing peeling front. Insets: configurations at beginning of contact and at later stage, as in (c).}
\label{figS2}
\end{center}

As in the derivation of the governing equations without contact (Methods), we shall express the variations in terms of $\delta r$ and $\delta\beta$. The third variation, $\delta z$, is not independent of the former two, and so the condition that the vertical positions of the shell at the point of contact and at the phialopore match must be incorporated via a Lagrange multiplier, $U$. \cite{julicher94} raised a related issue in the derivation of the shape equations for vesicles. The Lagrangian for the problem is therefore
\begin{equation}
\mathcal{L}=\mathcal{E}-2\pi U\int_C^Q{f_s\sin{\beta}\,\mathrm{d}\theta}, \label{eq:L5}
\end{equation}
where the prefactor has been introduced for mere convenience. We note the variation of Eq.~(\ref{eq:L5}),
\begin{equation}
\dfrac{\delta\mathcal{L}}{2\pi}=\dfrac{\delta\mathcal{E}}{2\pi}-\Bigl\llbracket U\tan{\beta}\,\delta r\Bigr\rrbracket_{C_+}^Q+Uf_s(C_+)\sin{\beta(C)}\,\delta C+U\int_C^Q{\Bigl\{f_s\kappa_s\sec^2{\beta}\,\delta r-f_s\sec{\beta}\,\delta\beta\Bigr\}\,\mathrm{d}\theta}.
\end{equation}
Next, expanding the condition $\beta(C_-)=\beta(C_+)$ of geometric continuity that we have already implicitly applied in the above, we note that
\begin{equation}
\delta\beta(C_-)+f_s(C_-)\kappa_s(C_-)\,\delta C=\delta\beta(C_+)+f_s(C_+)\kappa_s(C_+)\,\delta C. 
\end{equation}
Since the outer part of the shell can rotate freely with respect to the inner part at the points of contact, the variations $\delta\beta(C_\pm)$ and $\delta\beta(Q)$ are, by contrast, independent. This is not true of the variations $\delta r(C_\pm)$ and $\delta r(Q)$, however:
\begin{equation}
\delta r(Q) = \delta r(C_-)+f_s(C_-)\cos{\beta(C)}\,\delta C=\delta r(C_+)+f_s(C_+)\cos{\beta(C)}\,\delta C.
\end{equation}
Analogous expansions were used by \cite{seifert91} for discussing an adhesion problem for vesicles. Next, a straightforward calculation reveals that the governing equations (\ref{eq:Eb}) and (\ref{eq:ET}) remain unchanged if we define $T=-N_s\tan{\beta}+U\sec{\beta}/r$ for $C\leqslant \theta\leqslant Q$. For convenience, we adjoin the equation $\mathrm{d}z/\mathrm{d}s=f_s\sin{\beta}$ to the system (thereby fixing the degree of freedom of vertical translation). The system thus becomes a system of six first-order differential equations on two regions, with two unknown parameters (the contact position $C$ and the Lagrange multiplier~$U$). We thus have to impose fourteen boundary conditions:
\begin{subequations}
\begin{align}
r(0)&=0,&z(0)&=0,&\beta(0)&=0,&T(0)&=0,\\
r(Q)&= r(C),&z(Q)&= z(C),&N_s(Q)&=0,&M_s(Q)&=0,
\end{align}
as well as the continuity conditions at $\theta=C$,
\begin{align}
\llbracket\beta\rrbracket&=0,&\llbracket r\rrbracket&=0,&\llbracket z\rrbracket&=0,&\llbracket M_s\rrbracket&=0, 
\end{align}
and
\begin{align}
&r(C)\llbracket N_s\rrbracket\sec{\beta(C)}-U\tan{\beta(C)}=r(Q)N_s(Q)\sec{\beta(Q)}-U\tan{\beta(Q)},\\
&\llbracket T\rrbracket = -\llbracket N_s\rrbracket\tan{\beta(C)} + \dfrac{U\sec{\beta(C)}}{r(C)},\qquad\dfrac{\llbracket\mathcal{E}\rrbracket}{2\pi}=r(C)\bigl\llbracket f_sN_s\bigr\rrbracket+r(C)M_s(C)\llbracket f_s\kappa_s\rrbracket.
\end{align}
\end{subequations}
We also note that the conditions $r(Q)=r(C)$ and $z(C)=z(Q)$ do not take into account the finite, but small, thickness of the shell. A more detailed condition would require knowledge of the nature of the contact (and is anyway beyond the remit of a thin shell theory). 

We briefly explore shapes in the contact configuration in what follows. We start from a configuration where the posterior hemisphere has inverted, as in Fig.~\ref{fig:antpeeling}d, and advance the peeling front, but now without increasing the intrinsic circumferential stretch $f_\phi^0$ at the phialopore. As the peeling front advances, the circumferential stretch at the phialopore increases (Fig.~\ref{figS2}c) at constant $f_\phi^0$, showing how the phialopore is pushed open by the posterior hemisphere. The procession of the point of contact between the posterior and the phialopore along the inverted posterior speeds up with advancing peeling front position (Fig.~\ref{figS2}d) because the closer the point of contact is to the posterior, the more the latter resists the progression of the contact point because of the changing tangent angle.

The inset configurations in Fig.~\ref{figS2}c,d also suggest that, as the peeling front advances, the regime of contact at a point discussed here gives way to a second contact regime, where the contact is over a finite extent of the meridian of the shell. We do not pursue this further.

\subsection{Asymptotic Analysis of a Toy Problem}
Some analytic progress can be made and additional insight into the contact configuration can be gained by asymptotic analysis of a toy problem: two elastic spherical shells, an inner shell of radius $R_1$ and an outer, open shell of radius $R_2>R_1$, touch at the respective angular positions $\mathit{\Theta}_1$ and $\mathit{\Theta}_2<\mathit{\Theta}_1$ (Fig.~\ref{figS3}a), so that $R_2/R_1=\sin{\mathit{\Theta}_1}/\sin{\mathit{\Theta}_2}$. The intrinsic stretches and curvatures are those of the undeformed shells. For the remainder of this section, we non-dimensionalise distances with respect to the radius $R_1$ of the inner shell; stresses we non-dimensionalise with $Eh$. 

If the outer shell is moved relative to the inner shell by a distance $d$ (Fig.~\ref{figS3}b), the two shells deform in asymptotically small regions near the point of contact. This point of contact moves a distance $d\mathit{\Xi}$ down along the inner shell, determined by matching the displacements of the contact point and the forces exerted by one shell on the other. We assume in particular that the nature of the contact is such that the shells do not exert  torques on each other. Since we have non-dimensionalised distances with $R_1$, our asymptotic small parameter is
\begin{equation}
\varepsilon^2=\dfrac{1}{12(1-\nu^2)}\dfrac{h^2}{R_1^2}\ll 1.
\end{equation}
\begin{center}
\includegraphics{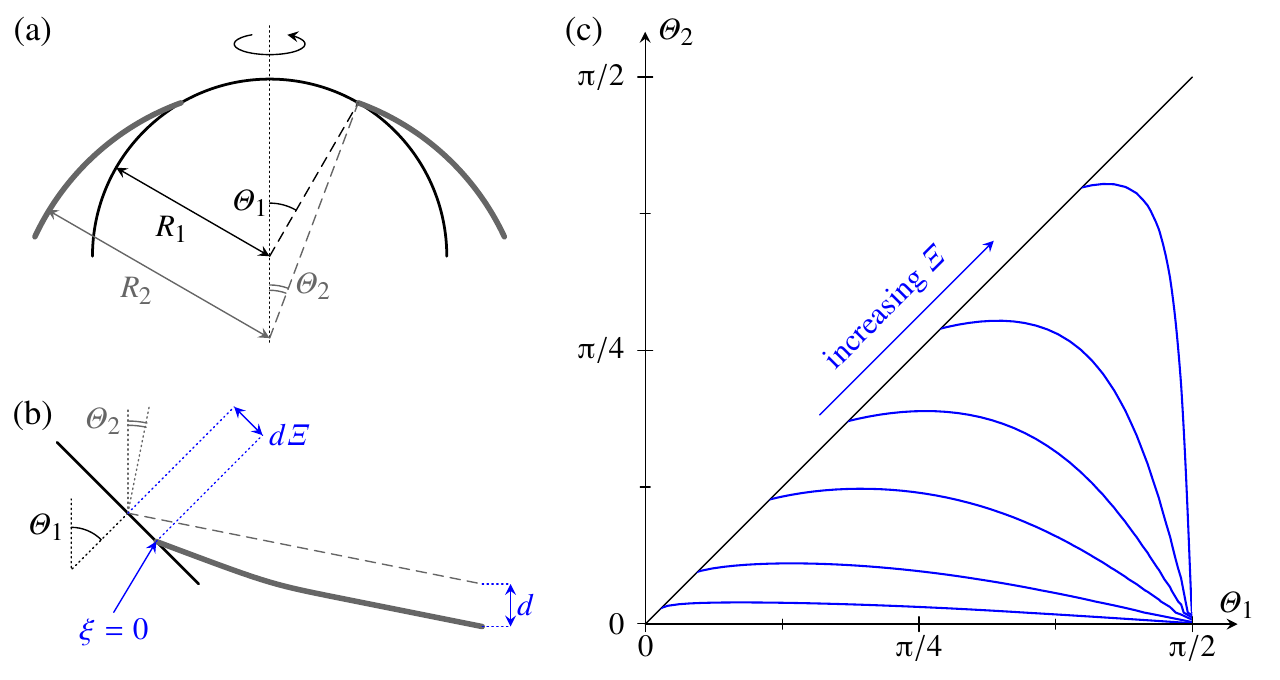}
\captionof{figure}{Asymptotic toy contact problem. (a) Two shells of radii $R_1$ and $R_2$ are in contact at angular positions $\mathit{\Theta}_1$ and $\mathit{\Theta}_2$, respectively. (b)~Relative motion of one shell with respect to the other by a distance $d$ induces deformations of the shell in an asymptotic inner layer of size~$\delta$, and causes the point of contact to move by a distance $d\mathit{\Xi}$ along the inner shell. (c)~Contours of $\mathit{\Xi}$ in the $(\mathit{\Theta}_1,\mathit{\Theta}_2)$ plane.}
\label{figS3}
\end{center}
The classical leading-order scalings for this problem are discussed by \cite{haas15}, for example: deformations are localised to asymptotic inner regions of width $\delta\sim\varepsilon^{1/2}$, in which deviations of the tangent angle from its equilibrium value are of order $d/\delta$, and we assume that $d\ll \delta$. We introduce an inner coordinate $\xi$, and write the polar angles as $\theta_1=\mathit{\Theta}_1+\delta\xi+\mathcal{O}(d)$, $\theta_2=\mathit{\Theta}_2+\delta \xi$. We thus expand
\begin{align}
\beta_1(\theta_1)&=\mathit{\Theta}_1+(d/\delta)b_1(\xi), &\beta_2(\theta_2)&=\mathit{\Theta}_2+(d/\delta)b_2(\xi).
\end{align}
Assuming that $\delta^2\ll d\ll\delta$, we then have the leading-order expansions
\begin{subequations}
\begin{align}
N_s^{(1)}&\stackrel{\hphantom{(\ast)}}{=}Eh\,\delta d\,\sigma_1(\xi), &N_s^{(2)}&\stackrel{\hphantom{(\ast)}}{=}Eh\,\delta d\,\sigma_2(\xi),\\
N_\phi^{(1)}&\stackrel{(\ast)}{=}Eh\,E_\phi^{(1)}+\nu N_s^{(1)}=Eh\,\delta\,a_1(\xi),&N_\phi^{(2)}&\stackrel{(\ast)}{=}Eh\,E_\phi^{(2)}+\nu N_s^{(2)}=Eh\,\delta\,a_2(\xi),
\end{align}
\end{subequations}
where $a_1,a_2$ are hoop strains. We note that the relations marked $(\ast)$ are only valid at leading order, where we may approximate $f_s\approx f_\phi\approx 1$. Let $F_r$ and $F_z$ denote the (suitably scaled) radial and vertical forces exerted by the outer shell on the inner shell. We obtain the leading-order force balances from the energy variation (\ref{eq:var2}): using dashes to denote differentiation with respect to $\xi$,
\begin{align}
&\sigma_1'\sin^2{\mathit{\Theta}_1}-b_1'''\cos{\mathit{\Theta}_1}\sin{\mathit{\Theta}_1}=F_z\delta(\xi),&&\sigma_1'\sin{\mathit{\Theta}_1}\cos{\mathit{\Theta}_1}-a_1+b_1'''\sin^2{\mathit{\Theta}_1}=F_r\delta(\xi).
\end{align}
This system is closed, at leading order, by the geometric relation $a_1'=-b_1$, as in \cite{haas15}. Eliminating $\sigma_1$, we obtain
\begin{equation}
b_1''''+b_1=\bigl(F_r-F_z\cot{\mathit{\Theta}_1}\bigr)\delta'(\xi). 
\end{equation}
The matching conditions $b_1\rightarrow 0$ as $\xi\rightarrow\pm\infty$ reduce the number of undetermined constants to four, which are determined by the jump conditions at the contact point $\xi=0$.

The asymptotic balance for the outer shell is of course the same, but we must remember that the system has been non-dimensionalised with the radius of the inner shell, for which reason a geometric factor arises in the equations. Thus
\begin{equation}
b_2''''+\left(\dfrac{\sin{\mathit{\Theta}_1}}{\sin{\mathit{\Theta}_2}}\right)^4b_2=0,
\end{equation}
with the matching condition $b_2\rightarrow 0$ as $\xi\rightarrow\infty$, leaving two boundary conditions to be imposed on this equation. Since the shells do not exert any moments on each other, $b_2'(0)=0$. The second condition is obtained from the force balance: the vertical force balance can be integrated once to yield
\begin{equation}
\sin{\mathit{\Theta}_1}\sin{\mathit{\Theta}_2}\left\{\sigma_2-\cot{\mathit{\Theta}_2}\left(\dfrac{\sin{\mathit{\Theta}_2}}{\sin{\mathit{\Theta}_1}}\right)^4b_2''\right\}=F_z. \label{eq:vert}
\end{equation}
Matching to the undeformed, unstressed shell as $\xi\rightarrow\infty$ implies $F_z=0$. The radial force boundary condition resulting from (\ref{eq:var2}) is
\begin{equation}
\sin{\mathit{\Theta}_1}\cos{\mathit{\Theta}_2}\left\{\sigma_2(0)+\tan{\mathit{\Theta}_2}\left(\dfrac{\sin{\mathit{\Theta}_2}}{\sin{\mathit{\Theta}_1}}\right)^4b_2''(0)\right\}=F_r, 
\end{equation}
which, upon imposing (\ref{eq:vert}), reduces to
\begin{equation}
b_2''(0)=\left(\dfrac{\sin{\mathit{\Theta}_1}}{\sin{\mathit{\Theta}_2}}\right)^3F_r. 
\end{equation}
Let $U_r^{(1)},U_z^{(1)}$ and $U_r^{(2)},U_z^{(2)}$ denote the respective (non-dimensional) displacements of the contact point $\xi=0$, scaled with $d$. Then
\begin{subequations}
\begin{align}
U_r^{(1)}&=\sin{\mathit{\Theta}_1}\int_0^\infty{b_1\,\mathrm{d}\xi}=-\dfrac{F_r}{2\sqrt{2}}\sin{\mathit{\Theta}_1},&U_z^{(1)}&=-\cos{\mathit{\Theta}_1}\int_0^\infty{b_1\,\mathrm{d}\xi}=\dfrac{F_r}{2\sqrt{2}}\cos{\mathit{\Theta}_1},\\
U_r^{(2)}&=\sin{\mathit{\Theta}_1}\int_0^\infty{b_2\,\mathrm{d}\xi}=-\sqrt{2}F_r\sin{\mathit{\Theta}_1},&U_z^{(2)}&=-\sin{\mathit{\Theta}_1}\cot{\mathit{\Theta}_2}\int_0^\infty{b_2\,\mathrm{d}\xi}=\sqrt{2}F_r\sin{\mathit{\Theta}_1}\cot{\mathit{\Theta}_2}.
\end{align}
\end{subequations}
In particular, these expressions once again contain additional geometric factors resulting from the non-dimensionalisation. 

The values of the two remaining undetermined constants, $F_r$ and $\mathit{\Xi}$, are finally obtained by imposing continuity of the displacement of the contact point, i.e.
\begin{align}
U_r^{(1)}+\mathit{\Xi}\cos{\mathit{\Theta}_1}&=U_r^{(2)},&U_z^{(1)}+\mathit{\Xi}\sin{\mathit{\Theta}_1}&=U_z^{(2)}+1.
\end{align}
Notice that arclength is computed here from the anterior pole of the shell to match the asymptotic setup of \cite{haas15}, and so the `vertical' axis is pointing downwards in Fig.~\ref{figS3}, giving rise to some sign changes. In particular, we obtain
\begin{equation}
\mathit{\Xi}=\dfrac{3\sin{\mathit{\Theta}_1}}{1+2\operatorname{cosec}{\mathit{\Theta}_2}\sin{\bigl(2\mathit{\Theta}_1-\mathit{\Theta}_2\bigr)}}.
\end{equation}
The contours of this expression are plotted in Fig.~\ref{figS3}c. The very non-linear nature of this expression illustrates that the contact geometry is quite intricate; in particular, $\mathit{\Theta}_2(\mathit{\Theta}_1)$ at fixed $\mathit{\Xi}$ is not a monotonic function, but, as expected (since it is easier for the the contact point to slide along the inner shell the more parallel it is to the axis of symmetry), at fixed~$\mathit{\Theta}_1$, $\mathit{\Xi}$ increases with $\mathit{\Theta}_2$.
\end{appendixbox}
\end{document}